\documentclass[12pt]{article}

\catcode`\@=11
\@addtoreset{equation}{section}

\global\arraycolsep=2pt 
\usepackage{mathtools,mathrsfs, amsbsy, amssymb, latexsym, amsfonts, amscd, amsmath, cite} 
\usepackage{graphicx,color}

\allowdisplaybreaks

\newcommand{\ga}{\gamma}

\newcommand{\la}{\lambda}
\newcommand{\ze}{\zeta}

\newcommand{\Tr}{{\rm Tr}}

\newcommand{\alg}[1]{\mathfrak{#1}}

\newcommand{\nln}{\nonumber\\}

\newcommand{\E}{\mathcal E}

\newcommand{\mL}{\mathcal L}

\newcommand{\txt}{\textrm}

\usepackage{mathrsfs,amsbsy,amssymb,latexsym,amsfonts,amsmath,cite}
\usepackage{graphicx,color}
\usepackage{mathtools}

\begin{document}

\begin{flushright}
\parbox{4cm}
{KUNS-2575}
\end{flushright}

\vspace*{1.5cm}

\begin{center}
{\Large \bf Lax pairs on Yang-Baxter deformed backgrounds}
\vspace*{1.5cm}\\
{\large Takashi Kameyama\footnote{E-mail:~kame@gauge.scphys.kyoto-u.ac.jp},
Hideki Kyono\footnote{E-mail:~h\_kyono@gauge.scphys.kyoto-u.ac.jp},
Jun-ichi Sakamoto\footnote{E-mail:~sakajun@gauge.scphys.kyoto-u.ac.jp},\\
and Kentaroh Yoshida\footnote{E-mail:~kyoshida@gauge.scphys.kyoto-u.ac.jp}} 
\end{center}

\vspace*{0.25cm}

\begin{center}
{\it Department of Physics, Kyoto University, \\ 
Kitashirakawa Oiwake-cho, Kyoto 606-8502, Japan} 
\end{center}

\vspace{2cm}

\begin{abstract}
We explicitly derive Lax pairs for string theories on Yang-Baxter deformed backgrounds, 
1) gravity duals for noncommutative gauge theories, 2) $\gamma$-deformations of S$^5$, 
3) Schr\"odinger spacetimes and 4) abelian twists of the global AdS$_5$\,. 
Then we can find out a concise derivation of Lax pairs based on simple replacement rules. 
Furthermore, each of the above deformations can be reinterpreted as 
a twisted periodic boundary conditions with the undeformed background by using the rules.
As another derivation, the Lax pair for gravity duals for noncommutative gauge theories is reproduced 
from the one for a $q$-deformed AdS$_5\times$S$^5$ by taking a scaling limit. 
\end{abstract}

\setcounter{footnote}{0}
\setcounter{page}{0}
\thispagestyle{empty}

\newpage

\tableofcontents

\section{Introduction}

The AdS/CFT correspondence is a fascinating subject in the study of string theory. 
The most famous one among a lot of variations is a duality between 10D type IIB string theory 
on the AdS$_5\times$S$^5$ background 
and the 4D $\mathcal{N} = 4$ super Yang-Mills theory at large $N$ limit \cite{M}.  
A great progress is that an integrable structure has been discovered behind this duality \cite{review}.
On the string-theory side, the Green-Schwarz string action on AdS$_5\times$S$^5$ is constructed 
as a 2D coset sigma model \cite{MT} and the $\mathbb{Z}_4$-grading of the supercoset 
ensures the classical integrability \cite{BPR} 
(For a big review of the AdS$_5\times$S$^5$ superstring, see \cite{AF-review}). 
Although the essential mechanism of the duality has not been fully understood yet,  
the integrability has played a crucial role in checking conjectured relations in the AdS/CFT. 

\medskip

It would be significant to consider integrable deformations of the AdS/CFT. 
It may shed light on a deeper structure behind gauge/gravity dualities beyond the conformal invariance. 
On the string-theory side, an influential way  is to employ the Yang-Baxter sigma model description \cite{Klimcik}.
This is a systematic way to study integrable deformations of 2D non-linear sigma models. 
By following this approach, an integrable deformation is specified by picking up a skew-symmetric classical $r$-matrix 
which satisfies the modified classical Yang-Baxter equation (mCYBE). 
The original argument \cite{Klimcik} was restricted to principal chiral models. 
It is generalized to symmetric cosets by Delduc-Magro-Vicedo \cite{DMV}\footnote{
For earlier arguments related to this generalization, see \cite{KYhybrid,KMY-QAA,KOY,Sch}.}.
Then they succeeded in constructing a $q$-deformed action of the AdS$_5\times$S$^5$ superstring \cite{DMV-string}.
This formulation is also based on the mCYBE. 

\medskip 

After that, it has been reformulated in \cite{KMY-JordanianAdSxS} 
based on the (non-modified) classical Yang-Baxter equation (CYBE), 
where the Lax pair and the kappa transformation should be reconstructed 
and this generalization is not so trivial. An advantage in comparison to the mCYBE case 
is that partial deformations of  AdS$_5\times$S$^5$ can be considered. 
This is because the zero map $R=0$ is allowed for the CYBE while not for the mCYBE. 
Furthermore, one can find many solutions of the CYBE. 
In fact, in a series of papers \cite{KMY-SUGRA, MY1, MY2, Sch-YB, YB1, YB2, Stijn}, 
many examples of (skew-symmetric) classical $r$-matrices have been identified with 
the well-known backgrounds such as $\ga$-deformations of S$^5$ \cite{LM,Frolov}, 
gravity duals for noncommutative (NC) gauge theories \cite{HI} and Schr\"odinger spacetimes \cite{10DSch}, 
in addition to new backgrounds \cite{KMY-SUGRA}. 
This identification may be called the gravity/CYBE correspondence \cite{MY1} (For a short summary, see \cite{CYBE}) 
and indicate that the moduli space of (a certain class of) solutions of type IIB supergravity 
can be described by the CYBE\footnote{ 
This is quit analogous to the bubbling scenario proposed by Lin-Lunin-Maldacena \cite{LLM}. 
Here the moduli space is described by droplet configurations in a free fermion system.}. 

\medskip 

In the recent, this correspondence has been generalized to integrable deformations of 4D Minkowski spacetime 
\cite{Minkowski}. In particular, (T-duals of) 4D (A)dS spaces are reproduced as Yang-Baxter deformations of 
the Minkowski spacetimes. Furthermore, this development has an intimate connection with kappa-Minkowski 
spacetime \cite{kappa} via preceding works e.g., \cite{r-dS}. 
For a recent argument with a gravity dual, see \cite{Stijn}. 

\medskip 

It is also remarkable that the gravity/CYBE correspondence seems to be valid beyond the integrability. 
There are many examples of non-integrable AdS/CFT correspondences.  
An example is the case of AdS$_5\times T^{1,1}$ \cite{KW}, for which the non-integrability 
has been shown by the existence of chaotic string solutions on $R \times T^{1,1}$ \cite{BZ,Penrose-T11}. 
Thus TsT transformations of $T^{1,1}$ \cite{LM,CO} are regarded as non-integrable deformations. 
However, these deformations can be described as Yang-Baxter deformations \cite{CMY-T11}. 
Hence the gravity/CYBE correspondence would be applicable to a much wider class 
of solutions of type IIB supergravity. 
 
\medskip

We will proceed to study Yang-Baxter deformations of the AdS$_5\times$S$^5$ superstring 
by focusing upon Lax pairs. The universal expression of Lax pair, without taking specific classical $r$-matrices 
and concrete coordinate systems, 
has already been presented in \cite{KMY-JordanianAdSxS}. 
Then classical $r$-matrices are identified and hence the associated Lax pairs are already obtained in an implicit way. 
However, explicit expressions of the Lax pairs have not been evaluated yet, 
while those are quite useful in studying classical solution by the use of the classical inverse scattering method. 
Thus, in this paper, we will derive explicit expressions of Lax pairs for string theories 
on popular examples of deformed backgrounds: 
1) gravity duals for NC gauge theories \cite{HI}, 
2) $\gamma$-deformations of S$^5$ \cite{LM,Frolov}, 3) Schr\"odinger spacetimes \cite{10DSch} 
and 4) abelian twists of the global AdS$_5$ \cite{MS}. 
Then we can find out a concise derivation of Lax pairs based on simple replacement rules. 
Furthermore, each of the above deformations can be reinterpreted as 
a twisted periodic boundary conditions with the undeformed background by using the rules. 
As another derivation, the Lax pair for gravity duals for NC gauge theories is reproduced 
from the one for a $q$-deformed AdS$_5\times$S$^5$ \cite{DMV-string,ABF} 
by taking a scaling limit introduced in \cite{ABF2}.

\medskip

This paper is organized as follows.
Section 2 gives a brief review of Yang-Baxter deformations of the  AdS$_5\times$S$^5$ superstring.
In section 3, we explicitly derive Lax pairs for gravity duals of NC gauge theories with two methods, 
(i) Yang-Baxter deformation (argued in \cite{MY2}) and 
(ii) a scaling limit of $q$-deformed AdS$_5\times$S$^5$ (introduced in \cite{ABF2}).
The resulting Lax pairs are identical under a unitary transformation.  
In section 4, we compute Lax pairs for $\gamma$-deformations of S$^5$ 
by evaluating the abstract expression given in \cite{MY1}. The resulting Lax pairs 
agree with Frolov's results \cite{Frolov} up to gauge transformations. 
Section 5 argues Lax pairs for Schr\"odinger spacetimes by evaluating the abstract forms given in \cite{Sch-YB}. 
In section 6, we derive Lax pairs for abelian twists of the global AdS$_5$ from the results of \cite{MY1}. 
Section 7 is devoted to conclusion and discussion.
In Appendix A, our convention and notation are summarized.
In Appendix B, we explicitly derive a Lax pair for a $q$-deformed AdS$_5\times$S$^5$ 
by evaluating the universal Lax pair \cite{DMV-string} with a coordinate system \cite{ABF}.

\section{Yang-Baxter deformations of string on AdS$_5\times$S$^5$} 

We shall give a brief review of Yang-Baxter deformations of the AdS$_5\times$S$^5$ superstring action 
based on the CYBE case \cite{KMY-JordanianAdSxS}\footnote{
For the mCYBE case \cite{DMV-string}, see Appendix B.}.

\medskip

The deformed classical action of the AdS$_5\times$S$^5$ superstring is given by
\begin{eqnarray}
S=-\frac{\sqrt{\lambda_{\rm c}}}{4}\int_{-\infty}^\infty d\tau\int_0^{2\pi}d\sigma\,
(\ga^{\alpha\beta}-\epsilon^{\alpha\beta}){\rm STr}\Bigl[A_\alpha\, d\circ\frac{1}{1-\eta R_g\circ d}(A_\beta)\Bigr]\,, 
\label{YBsM}
\end{eqnarray}
where the left-invariant one-form $A_\alpha$ is defined as
\begin{eqnarray}
A_\alpha\equiv g^{-1}\partial_\alpha g\,,\quad\quad g\in SU(2,2|4) 
\end{eqnarray}
with the world-sheet index $\alpha=(\tau,\sigma)$\,. 
Here the conformal gauge is supposed and the world-sheet 
metric is taken as $\ga^{\alpha\beta}={\rm diag}(-1,+1)$\,. Hence there is no coupling of the dilaton to 
the world-sheet scalar curvature. 
The anti-symmetric tensor $\epsilon^{\alpha\beta}$ is normalized as $\epsilon^{\tau\sigma}=+1$\,. 
The constant $\lambda_{\rm c}$ is the 't Hooft coupling.  
Note that $\eta$ is a deformation parameter and hence the undeformed action \cite{MT} is reproduced when $\eta=0$\,.

\medskip

A key ingredient in our analysis is the operator $R_g$ defined as
\begin{eqnarray}
R_g(X)\equiv g^{-1}R(gXg^{-1})g\,, \quad\quad X\in \alg{su}(2,2|4)\,,
\label{R}
\end{eqnarray}
where a linear $R$-operator $R:\alg{su}(2,2|4)\to \alg{su}(2,2|4)$ 
is a solution of the classical Yang-Baxter equation (CYBE)\footnote{In the original work \cite{KMY-JordanianAdSxS}, 
a wider class of $R$-operators whose image is given by $\mathfrak{gl}(4|4)$ has been proposed.  
The $\mathfrak{gl}(4|4)$ image is restricted on $\alg{su}(2,2|4)$ in essential under the coset projection $d$ 
as pointed out in \cite{Stijn}. 
We will concentrate here on a restricted class in which the image is $\alg{su}(2,2|4)$ from the beginning, 
so as to deal with pre-projected quantities like the deformed current $J$ itself, without introducing extra generators. 
For general cases argued in \cite{KMY-SUGRA,YB1}, a more detailed study would be necessary. },
\begin{eqnarray}
[R(X),R(Y)]-R([R(X),Y]+[X,R(Y)])=0\,. \label{CYBE}
\end{eqnarray}
This $R$-operator is related to a skew-symmetric classical $r$-matrix in the tensorial notation 
through the following supertrace operation on the second site: 
\begin{eqnarray}
R(X)=\text{STr}_2[r(1\otimes X)]=\sum_i\left( a_i\,\text{STr}[b_iX]-b_i\,\text{STr}[a_iX]\right)\,,
\label{R-r}
\end{eqnarray}
where the classical $r$-matrix is represented by 
\begin{eqnarray}
r=\sum_ia_i \wedge  b_i\equiv\sum_i \left(a_i \otimes b_i - b_i \otimes a_i\right)\qquad 
\txt{with}\qquad a_i,b_i\in\mathfrak{su}(2,2|4)\,.
\end{eqnarray}

\medskip

The projection operator $d$ is defined as 
\begin{eqnarray}
d &\equiv& P_1+2P_2-P_3\,, 
\label{op-d}
\end{eqnarray}
where $P_i\,(i=0,1,2,3)$ are projections to the $\mathbb{Z}_4$-graded components of $\mathfrak{su}(2,2|4)$\,.
In particular, $P_0(\mathfrak{su}(2,2|4))$ is a local symmetry of the classical action, 
$\mathfrak{so}(1,4)\oplus\mathfrak{so}(5)$\,.
Note that the numerical coefficients are fixed by requiring the kappa-symmetry
\cite{KMY-JordanianAdSxS}.

\medskip

It is convenient to introduce the light-cone expression of $A_\alpha$ like 
\begin{eqnarray}
A_{\pm} \equiv A_\tau\pm A_\sigma\,,
\end{eqnarray}
when we will study Lax pair in the following sections.

\subsubsection*{The bosonic part of the Lagrangian}

Our aim here is to explicitly derive Lax pairs for the bosonic part of deformed actions.
Hence it is convenient to rewrite the bosonic part of the deformed Lagrangian (\ref{YBsM}) as
\begin{eqnarray}
L=\frac{\sqrt{\lambda_{\rm c}}}{2}\,\text{STr}(A_-\,P_2(J_+))\,,
\label{L}
\end{eqnarray}
where $J_\pm$ is a deformed current defined as 
\begin{eqnarray}
J_\pm \equiv \frac{1}{1\mp 2\eta R_g \circ P_2}\, A_\pm\,.
\label{J-CYBE}
\end{eqnarray}
Note here that the factor 2 in front of $\eta$ comes from the projection operator $d$ given in (\ref{op-d})\,. 
By solving the following equations,
\begin{eqnarray}
\left(1 \mp 2\eta R_g \circ P_2 \right) J_{\pm} = A_{\pm}\,, 
\label{PJ-CYBE}
\end{eqnarray}
the deformed current $J_\pm$ is determined\footnote{In order to derive 
the metric and NS-NS two-form, it is enough to determine $P_2(J_{\pm})$ 
by solving the projected conditions 
\[
\left(1\mp 2\eta P_2 \circ  R_g \right)P_2(J_\pm) = P_2(A_\pm)\,, 
\] 
as done in a series of the previous papers \cite{KMY-SUGRA, MY1, MY2, Sch-YB, YB1, YB2, Stijn}. 
However, it is necessary here to determine $J_{\pm}$ themselves so as to evaluate the form of Lax pair.}.  
Then the metric and NS-NS two-form are evaluated from the symmetric and skew-symmetric parts  
regarding the world-sheet coordinates in (\ref{L}), respectively.

\medskip

Taking a variation of the Lagrangian (\ref{L}), the equation of motion is obtained as 
\begin{eqnarray}
\mathcal{E}\equiv\partial_+P_2(J_-)+\partial_-P_2(J_+)+[J_+,P_2(J_-)]+[J_-,P_2(J_+)]
=0\,.\label{eom}
\end{eqnarray}
By definition, the undeformed current $A_\pm$ satisfies the flatness condition,
\begin{eqnarray}
\mathcal{Z}\equiv\partial_+A_--\partial_-A_++[A_+,A_-]
=0\,.\label{flat}
\end{eqnarray}
Then, in terms of the deformed current $J_\pm$\,, this condition can be rewritten as
\begin{eqnarray}
\partial_+J_--\partial_-J_++[J_+,J_-]+2\eta\, R_g(\E)+4\eta^2\,{\rm CYBE}_g(P_2(J_+),P_2(J_-)) =0\,,
\label{flat2}
\end{eqnarray}
where we have introduced a new quantity defined as 
\begin{eqnarray}
{\rm CYBE}_g(X,Y) \equiv [R_g(X),R_g(Y)]-R_g([R_g(X),Y]+[X,R_g(Y)])\,.
\end{eqnarray}
Note that ${\rm CYBE}_g(X,Y)$  vanishes if the $R$-operator satisfies the CYBE in (\ref{CYBE}).
The relation (\ref{flat2}) means that $J_\pm$ also satisfies the flatness condition with the equation of motion $\E=0$\,. 
That is, $J_{\pm}$ satisfies the flatness condition only on the on-shell, 
while $A_{\pm}$ do even on the off-shell.

\medskip

It is helpful to decompose $J_{\pm}$ with the projection operators $P_0$ and $P_2$ like 
\begin{eqnarray}
 J_{\pm} = P_0(J_{\pm}) + P_2(J_{\pm}) \equiv J_{\pm}^{(0)} + J_{\pm}^{(2)}\,, 
\end{eqnarray}
where we have used the completeness condition $P_0 + P_2 =1$\,. For the concrete expressions of the projection operators, 
see Appendix A. 
Then the equation of motion (\ref{eom}) can be rewritten into the following form: 
\begin{eqnarray}
\mathcal{E} = \partial_+J^{(2)}_-+\partial_-J^{(2)}_++[J^{(0)}_+,J^{(2)}_-]+[J^{(0)}_-,J^{(2)}_+]=0\,.\label{eom2}
\end{eqnarray}
The flatness condition (\ref{flat}) can also be rewritten in a similar way: 
\begin{eqnarray}
\mathcal{Z} = P_0(\mathcal{Z}) + P_2(\mathcal{Z})=0\,. 
\end{eqnarray}
With the help of the linear independence of the grade 0 and grade 2 parts, one can obtain the following two conditions: 
\begin{eqnarray}
P_0(\mathcal{Z})=\partial_+J^{(0)}_--\partial_-J^{(0)}_++[J^{(0)}_+,J^{(0)}_-]+[J^{(2)}_+,J^{(2)}_-]
+2\eta\,P_0( R_g(\E))=0\,,\nonumber \\ 
P_2(\mathcal{Z})=\partial_+J^{(2)}_--\partial_-J^{(2)}_++[J^{(0)}_+,J^{(2)}_-]+[J^{(2)}_+,J^{(0)}_-]
+2\eta\,P_2( R_g(\E))=0\,. \label{z2}
\end{eqnarray}
Note here that the terms proportional to $\eta$ vanish on the on-shell, i.e., $\mathcal{E}=0$\,. 

\medskip 

Then the three conditions in (\ref{eom2}) and (\ref{z2}) can be recast into the following set of the equations 
$\mathcal{C}_i=0~(i=1,2,3)$~:
\begin{eqnarray}
 \mathcal{C}_1 &\equiv & \partial_-J_+^{(2)}-[J_+^{(2)},J_-^{(0)}]\,,  \nonumber\\
 \mathcal{C}_2 &\equiv & \partial_+J_-^{(2)}+[J_+^{(0)},J_-^{(2)}]\,, \nonumber\\ 
 \mathcal{C}_3 &\equiv & \partial_+J_-^{(0)}-\partial_-J_+^{(0)}+[J_+^{(0)},J_-^{(0)}] +[J_+^{(2)},J_-^{(2)}]\,. 
\end{eqnarray}
Namely, $\mathcal{C}_i =0~(i=1,2,3)$ are satisfied on the on-shell and are equivalent to the equation of motion 
(\ref{eom}) and the flatness condition (\ref{flat}). 

\subsubsection*{Lax pair}

Finally, a Lax pair for the deformed action is given by
\begin{eqnarray}
\mL_\pm=J^{(0)}_\pm+\la^{\pm1}J^{(2)}_\pm 
\label{lax0}
\end{eqnarray}
with  a spectral parameter $\lambda\in\mathbb{C}$\footnote{
Please do not confuse the spectral parameter $\lambda$ with the 't Hooft coupling $\lambda_{\rm c}$!}. 
Note that the existence of the Lax pair (\ref{lax0}) is 
based on the $\mathbb{Z}_2$-grading of AdS$_5\times$S$^5$\,.  

\medskip 

As a matter of course, the flatness condition of $\mathcal{L}_{\pm}$ 
\begin{eqnarray}
0&=&\partial_+\mL_--\partial_-\mL_+ + [\mL_+,\mL_-] \label{L-flat}
\end{eqnarray}
is equivalent to the equation of motion $\mathcal{E}=0$ [in (\ref{eom})] 
and the flatness condition $\mathcal{Z}=0$ [in (\ref{flat})]\,. 
In order to confirm the equivalence, it is helpful to notice that 
the right-hand side of (\ref{L-flat}) can rewritten in terms of $\mathcal{C}_i$ as follows: 
\begin{eqnarray}
\partial_+\mL_--\partial_-\mL_+ + [\mL_+,\mL_-] = -\la\, \mathcal{C}_1+\frac{1}{\la}\,\mathcal{C}_2 + \mathcal{C}_3\,.
\end{eqnarray}
Thus we have shown the equivalence. 

\medskip 

In the following sections, we will evaluate explicit forms of the Lax pair (\ref{lax0}) 
for some examples of classical $r$-matrices.

\section{Lax pairs for gravity duals of NC gauge theories}

In this section, let us study Lax pairs for gravity duals of noncommutative (NC) gauge theories 
from the viewpoint of Yang-Baxter deformations. The integrability of this background 
was recently shown in \cite{MY2} in the sense of the kinematical integrability. 
The Lax pair was implicitly derived in \cite{MY2}, but the explicit expression has not been computed yet.

\medskip 
 
First of all, we evaluate explicit forms of Lax pairs with classical $r$-matrices in \cite{MY2}. 
The resulting Lax pairs depend on two deformation parameters.  
Next, one may consider another derivation for a special one-parameter case. 
Then the associated Lax pair can also be reproduced 
by taking a scaling limit \cite{ABF2} of the one for a $q$-deformed AdS$_5$ \cite{DMV-string,ABF}.

\subsection{Lax pairs from Yang-Baxter deformations}

In the context of Yang-Baxter deformations, abelian Jordanian classical $r$-matrices \cite{MY2} 
\begin{eqnarray}
r=c_1\,p_2\wedge p_3+c_2\,p_0\wedge p_1 \label{r-1}
\end{eqnarray}
are associated with gravity duals of NC gauge theories \cite{HI}. 
The classical $r$-matrices in (\ref{r-1}) consist of the translation generators $p_\mu$ 
in $\mathfrak{su}(2,2)$ (For our convention, see Appendix A). 
Then the deformation parameters $c_1$ and $c_2$ are related to 
magnetic and electric NS-NS two-forms in the gravity solutions \cite{HI}, respectively.  

\medskip 

Because the square of the associated $R$-operator vanishes due to the properties of $p_{\mu}$'s, 
the classical $r$-matrices (\ref{r-1}) are called Jordanian type. 
The classical $r$-matrices do not contain any generators in $\mathfrak{su}(4)$\,, 
and hence only the AdS$_5$ part is deformed. Therefore we will concentrate on only the AdS$_5$ part below.

\subsubsection*{The deformed metric and NS-NS two-form}

To derive the metric and NS-NS two-form from the Lagrangian (\ref{L})\,, 
let us introduce a coordinate system through a parametrization of an $SU(2,2)$ element as follows: 
\begin{eqnarray}
g_a(\tau,\sigma)=\exp\Bigl[p_0\, x^0 + p_1\, x^1 + p_2\, x^2 + p_3\, x^3\Bigr]\,
\exp\left[\gamma^a_5\, \frac{1}{2}\log z \right] \quad \in SU(2,2)\,. 
\end{eqnarray}
By solving the relation in (\ref{PJ-CYBE})\,, the deformed current $J_\alpha$ is explicitly determined as
\begin{eqnarray}
J_\pm&=&\frac{z}{z^4-4{c_2}^2\eta^2}\Bigl[
(z^2\partial_{\pm}x^0\pm2c_2\eta\partial_{\pm}x^1)\,p_0+(z^2\partial_{\pm}x^1\pm2c_2\eta\partial_{\pm}x^0)\,p_1
\Bigr] \nonumber \\
&&+\frac{z}{z^4+4{c_1}^2\eta^2}\Bigl[(z^2\partial_{\pm}x^2\pm2c_1\eta\partial_{\pm}x^3)\,p_2
+(z^2\partial_{\pm}x^3\mp2c_1\eta\partial_{\pm}x^2)\,p_3 \Bigr] \nonumber \\
&&+\frac{1}{2z}\partial_{\pm}z\,\gamma^a_5\,. \label{dc}
\end{eqnarray}
Then the resulting metric and NS-NS two-form are given by
\begin{eqnarray}
ds^2&=&\frac{z^2[-(dx^0)^2+(dx^1)^2]}{z^4-4c_2^2\eta^2}
+\frac{z^2[(dx^2)^2+(dx^3)^2]}{z^4+4c_1^2\eta^2}+\frac{dz^2}{z^2}\,,\nonumber \\
B&=&-\frac{2c_2\eta}{z^4-4c_2^2\eta^2}dx^0\wedge dx^1
+\frac{2c_1\eta}{z^4+4c_1^2\eta^2}dx^2\wedge dx^3\,.\label{MR}
\end{eqnarray} 
This result exactly agrees with the gravity duals of NC gauge theories \cite{HI}. 
When $c_1=c_2=0$\,, the Poincar\'e AdS$_5$ is reproduced.

\subsubsection*{Lax pair}

Let us derive the associated Lax pairs. Now $\mL^{\txt{NC}}_\pm$ are explicitly evaluated as   
\begin{eqnarray}
\mL^{\text{NC}}_\pm&=&\frac{z}{z^4-4c_2^2\eta^2}\Bigl[(z^2\partial_\pm x^0\pm2c_2\eta\partial_\pm x^1)
\Bigl(\frac{\la^{\pm1}}{2}\ga^a_0-n^a_{05}\Bigr) \nonumber \\ 
&& \qquad \qquad \quad +(z^2\partial_\pm x^1\pm2c_2\eta\partial_\pm x^0)
\Bigl(\frac{\la^{\pm1}}{2}\ga^a_1-n^a_{15}\Bigr)\Bigr]\nonumber \\
&& +\frac{z}{z^4+4c_1^2\eta^2}\Bigl[(z^2\partial_\pm x^2\pm2c_1\eta\partial_\pm x^3)
\Bigl(\frac{\la^{\pm1}}{2}\ga^a_2-n^a_{25}\Bigr) \nonumber \\ 
&& \qquad \qquad \quad~~ +(z^2\partial_\pm x^3\mp2c_1\eta\partial_\pm x^2)
\Bigl(\frac{\la^{\pm1}}{2}\ga^a_3-n^a_{35}\Bigr)\Bigr] 
+\frac{\la^{\pm1}\partial_\pm z}{2z}\ga^a_5\,.
\label{MR-lax}
\end{eqnarray}
In the undeformed limit $c_1,\,c_2\to0$\,,  the above expressions are reduced to
\begin{eqnarray}
\mL_\pm^{{\rm PAdS}_5} &=&\frac{\partial_\pm x^\mu }{z}
\Bigl(\frac{\la^{\pm 1}}{2}\ga^a_\mu-n^a_{\mu5}\Bigr)
+\frac{\la^{\pm1}\partial_\pm z}{2z}\ga^a_5\,. \label{P-lax}
\end{eqnarray}
This is nothing but a Lax pair for the Poincar\'e AdS$_5$\,.

\subsubsection*{Another derivation of Lax pair}

It would be of good significance to describe another derivation of Lax pair (\ref{MR-lax})\,. 

\medskip 

The undeformed current is now given by 
\begin{eqnarray}
A_\pm&=&\frac{1}{z}\partial_\pm x^\mu p_\mu+\frac{1}{2z}\partial_\pm z\,\gamma^a_5\,. \label{udc}
\end{eqnarray}
Then, by comparing the deformed current (\ref{dc}) with the undeformed one (\ref{udc})\,,  
the deformation under our consideration can be reinterpreted as the following replacement rules:
\begin{eqnarray}
\frac{1}{z^2}\,\partial_\pm x^0&& ~~\longrightarrow~~\frac{z^2}{z^4-4{c_2}^2\eta^2}
\left[\partial_\pm x^0 \pm \frac{2c_2\eta}{z^2}\partial_\pm x^1\right]\,,\nonumber\\
\frac{1}{z^2}\,\partial_\pm x^1&& ~~\longrightarrow~~ \frac{z^2}{z^4-4{c_2}^2\eta^2}
\left[\partial_\pm x^1\pm \frac{2c_2\eta}{z^2}\partial_\pm x^0\right]\,,\nonumber\\
\frac{1}{z^2}\,\partial_\pm x^2 && ~~\longrightarrow~~ \frac{z^2}{z^4+4{c_1}^2\eta^2}
\left[\partial_\pm x^2\pm \frac{2c_1\eta}{z^2}\partial_\pm x^3\right]\,,\nonumber\\
\frac{1}{z^2}\,\partial_\pm x^3&& ~~\longrightarrow~~ \frac{z^2}{z^4+4{c_1}^2\eta^2}
\left[\partial_\pm x^3\mp \frac{2c_1\eta}{z^2}\partial_\pm x^2\right]\,. \label{rule-1}
\end{eqnarray}
The above concise rules give rise to another simple derivation of the Lax pair (\ref{MR-lax}). 
By applying the rules to the undeformed Lax pair (\ref{P-lax})\,,  
the desired one (\ref{MR-lax}) can be reproduced. This derivation is quite similar to 
Frolov's construction of Lax pair for string on the $\gamma$-deformed S$^5$ \cite{Frolov}.

\subsubsection*{Twisted periodic boundary condition}
 
In fact, due to the  rule (\ref{rule-1})\,, the deformation can be regarded 
as a twisted periodic boundary condition with the undeformed AdS$_5\times$S$^5$\,, 
as argued in \cite{Frolov}. 

\medskip 

For simplicity, suppose $c_1 \neq 0$ and $c_2 =0$\,. The analysis for the case with $c_2 \neq 0$ is quite similar, 
though there is a subtlety for the signature of the metric (For the detail, see \cite{HI}). 

\medskip 

After performing the Yang-Baxter deformation (equivalently the associated TsT transformation)\,, 
the original coordinates $\tilde{x}^2$ and $\tilde{x}^3$ for the undeformed AdS$_5\times$S$^5$ 
are mapped to $x^2$ and $x^3$\,. Then the relations are given by 
\begin{eqnarray}
 \frac{1}{z^2}\,\partial_\pm \tilde{x}^2 &=& \frac{z^2}{z^4+4{c_1}^2\eta^2}
\left[\partial_\pm x^2\pm \frac{2c_1\eta}{z^2}\partial_\pm x^3\right]\,,\nonumber\\
\frac{1}{z^2}\,\partial_\pm \tilde{x}^3 &=& \frac{z^2}{z^4+4{c_1}^2\eta^2}
\left[\partial_\pm x^3\mp \frac{2c_1\eta}{z^2}\partial_\pm x^2\right]\,. 
\end{eqnarray}
These relation indicate the following equivalence of Noether currents 
\begin{eqnarray}
\tilde{P}_2^{\alpha} = P_2^{\alpha}\,, \qquad \tilde{P}_3^{\alpha} = P^{\alpha}_3\,, \label{rel}
\end{eqnarray}
where $P_i^{\alpha}$ and $\tilde{P}^{\alpha}_i$ $(i=2,3)$ 
are conserved currents associated with translation invariance for $x^i$ and $\tilde{x}^i$ directions, respectively. 
The $\tau$-component of the relations means that the momentum $p_i \equiv P_i^{\tau}$ 
is identical to $\tilde{p}_i \equiv \tilde{P}_i^{\tau}$\,, namely $p_i = \tilde{p}_i$\,.
Then, evaluating the $\sigma$-component of (\ref{rel}) leads to the relations:  
\begin{eqnarray}
\partial_{\sigma}\tilde{x}^2 = \partial_{\sigma}x^2 + \frac{2c_1\eta}{\sqrt{\lambda_{\rm c}}}\,p_3\,, 
\qquad 
\partial_{\sigma}\tilde{x}^3 = \partial_{\sigma}x^3 - \frac{2c_1\eta}{\sqrt{\lambda_{\rm c}}}\,p_2\,. 
\end{eqnarray}
Finally, by integrating these expressions, one can figure out that 
the deformed background with the usual periodic boundary condition 
is equivalent to the undeformed AdS$_5\times$S$^5$ with a twisted periodic boundary condition: 
\begin{eqnarray}
\tilde{x}^2 (\sigma=2\pi) = \tilde{x}^2(\sigma=0) + \frac{2c_1\eta}{\sqrt{\lambda_{\rm c}}}\,P_3\,, 
\qquad 
\tilde{x}^3 (\sigma=2\pi) = \tilde{x}^3(\sigma=0) - \frac{2c_1\eta}{\sqrt{\lambda_{\rm c}}}\,P_2\,.
\end{eqnarray}
Here $P_i$ are Noether charges for translation invariance in the $x^i$ directions. 

\medskip 

Thus the Yang-Baxter deformation with the classical $r$-matrix (\ref{r-1}) can be reinterpreted 
as a twisted periodic boundary condition with the usual AdS$_5\times$S$^5$\,.

\subsection{A scaling limit of a Lax pair for a $q$-deformed AdS$_5$}

In Section 3.1, we have derived the Lax pair (\ref{MR-lax}) 
as Yang-Baxter deformations of AdS$_5$\,. Here we shall reproduce it  
as a scaling limit of a Lax pair for a $q$-deformed AdS$_5$\,.

\subsubsection*{A scaling limit of a $q$-deformed AdS$_5\times$S$^5$}

We first give a short review of a scaling limit of a $q$-deformed AdS$_5\times$S$^5$ \cite{ABF2}. 
In this limit, the metric and NS-NS two-form in (\ref{MR}) can be reproduced.

\medskip

The starting point is the $q$-deformed metric and NS-NS two-form, 
\begin{eqnarray}
ds_{\txt{AdS}_5}^2&=& 
\sqrt{1+\varkappa^2}\,\biggl[\frac{1}{1-\varkappa^2\sinh^2\rho}
\left(-\cosh ^2\rho\, dt^2+d\rho^2\right)\nonumber \\ 
&&\hspace{1.5cm}+
\frac{\sinh^2\rho}{1+\varkappa^2\sin^2\zeta\sinh^4\rho}
\left[d\zeta^2+\cos ^2\zeta\,  (d\psi_1)^2\right] + \sinh ^2\rho \sin ^2\zeta\,( d\psi_2)^2\biggr]\,, 
\nonumber \\
B_{\txt{AdS}_5}&=&\varkappa\sqrt{1+\varkappa^2}\, 
\frac{\sinh ^4\rho\sin 2 \zeta}{1+\varkappa^2\sin^2\zeta\sinh^4\rho} \,d\psi_1\wedge d\zeta\,. 
\end{eqnarray}
Let us next rescale the coordinates as follows: 
\begin{eqnarray} 
&& t=\sqrt{\varkappa}\,x^0\,,\quad
\psi_1=\frac{\sqrt{\varkappa}}{\cos\zeta_0}\,x^2\,,\quad\psi_2=\frac{\sqrt{\varkappa}}{\sin\zeta_0}\,x^1\,, 
\nonumber\\
&&\zeta=\zeta_0+\sqrt{\varkappa}\,x^3\,,\qquad\rho=\txt{arcsinh}\left[\frac{1}{\sqrt{\varkappa}\,z}\right]\,. 
\label{scaling}
\end{eqnarray}
Here new coordinates $x^0\,,x^1\,,x^2\,,x^3\,,z$ and a real constant $\zeta_0$ 
have been introduced.

\medskip 

After taking the $\varkappa\to0$ limit, the resulting metric and NS-NS two-form are given by  
\begin{eqnarray}
ds^2&=&
 \frac{-(dx^0)^2+(dx^1)^2}{z^2}+\frac{z^2[(dx^2)^2+(dx^3)^2]}{z^4+\sin\zeta_0^2}+\frac{dz^2}{ z^2}\,, 
\nonumber \\
B&=&\frac{\sin\zeta_0}{z^4+\sin\zeta_0^2}\,dx^2\wedge dx^3\,.
\end{eqnarray}
This result exactly agrees with a one-parameter case of (\ref{MR}) through the identification 
\begin{eqnarray}
2c_1\,\eta=\sin\zeta_0\,, \qquad 2c_2\,\eta=0\,.
\label{c-eta}
\end{eqnarray}
For the S$^5$ part, this limit is nothing but the undeformed limit.

\subsubsection*{Lax pair -- the third derivation}

Next, we will derive the Lax pair (\ref{MR-lax}) by taking the scaling limit 
of a Lax pair for the $q$-deformed AdS$_5$\,. This is the third derivation. 

\medskip 

The Lax pair for the $q$-deformed AdS$_5\times$S$^5$ was originally constructed in \cite{DMV-string}. 
With a coordinate system \cite{ABF}, the Lax pair can be evaluated explicitly, 
as shown in Appendix B.  
The remaining task is to take the scaling limit of the Lax pair (\ref{q-lax})\,.

\medskip

The first is to rewrite the Lax pair (\ref{q-lax}) in terms of the coordinates (\ref{scaling}) 
with (\ref{c-eta})\,. 
For later convenience, the spectral parameter should be flipped as $\lambda \to-\lambda$\,.
Then, taking the $\varkappa\to0$ limit leads to the following expression:  
\begin{eqnarray}
\widetilde{\mL}_\pm&=&\frac{\partial_\pm x^0}{z}\left[-\frac{\la^{\pm1}}{2}i\ga_5^a+i n_{15}^a\right]
+\frac{\partial_\pm x^1}{z}\left[-i\frac{\la^{\pm1}}{2}\ga_0^a-in_{01}^a\right] \nonumber \\
&&+\frac{z(z^2\partial_\pm x^2+\eta\partial_\pm x^3)}{z^4+4c_1^2\eta^2}
\left[\frac{\la^{\pm1}}{2}\ga_2^a-n_{12}^a\right]\nonumber \\
&&+\frac{z(z^2\partial_\pm x^3-\eta\partial_\pm x^2)}{z^4+4c_1^2\eta^2}
\left[\frac{\la^{\pm1}}{2}\ga_3^a-n_{13}^a\right]
-\frac{\la^{\pm1}\partial_\pm z}{2z}\ga_1^a\,.\label{q-scal-lax}
\end{eqnarray}
In order to see that the result (\ref{q-scal-lax}) is identical to the Lax pair (\ref{MR-lax})\,, 
it is necessary to perform a unitary transformation like  
\begin{eqnarray}
\widetilde{\mL}_{\pm}
~~\longrightarrow~~ 
\mathcal{U}\,\widetilde{\mL}_{\pm}\,\mathcal{U}^{-1}\,, \qquad \mathcal{U} \equiv
\begin{pmatrix}
\;U~&~0\\
0&1\\
\end{pmatrix}\,,
\qquad U \equiv 
\begin{pmatrix}
\;1~&~i~&-i&1\\
i&1&-1&i\\
-i&1&1&i\\
-1&i&i&1\\
\end{pmatrix}\,.
\end{eqnarray}
After that, the transformed Lax pair sgrees with the one (\ref{MR-lax})\,, namely, 
\begin{eqnarray}
\mL^{{\rm NC}}_{\pm}=\mathcal{U}\,\widetilde{\mL}_{\pm}\,\mathcal{U}^{-1}\, 
\qquad \mbox{when}~~c_2=0\,.
\end{eqnarray}
Thus the scaling limit works well at the level of Lax pair. 

\medskip 

It would be nice to consider this relation at the level of classical $r$-matrix. 
One may interpret the scaling limit as a rescaling of Drinfeld-Jimbo type classical $r$-matrix \cite{DJ}.

\section{Lax pairs for $\gamma$-deformations of S$^5$}

In this section, we shall study Yang-Baxter deformations with classical $r$-matrices 
corresponding to $\gamma$-deformations of S$^5$\,. 
Concretely speaking, the associated Lax pairs are computed explicitly. 
The resulting expressions nicely agree with the Lax pairs obtained 
via TsT transformations of S$^5$ \cite{Frolov}. 
We will omit the AdS$_5$ part in the following. 

\medskip

Let us consider abelian classical $r$-matrices, which have been found in \cite{MY1}\,,  
\begin{eqnarray}
r=\mu_3\,h_4\wedge h_5+\mu_1\,h_5\wedge h_6+\mu_2\,h_6\wedge h_4\,. \label{abelian}
\end{eqnarray}
Here $h_4$\,, $h_5$ and $h_6$ are the three Cartan generators 
in $\mathfrak{su}(4)$\,, and $\mu_i$~($i=1,2,3$) are deformation parameters. 
For our convention of the generators, see Appendix A. 
The $r$-matrices (\ref{abelian}) deform only S$^5$ and correspond to $\gamma$-deformations of S$^5$\,.

\subsubsection*{The deformed metric and NS-NS two-form}

It is helpful to use the following representative of a group element of $SU(4)$\,,  
\begin{eqnarray}
g_s(\tau,\sigma) = \exp \left[\frac{i}{2}(\phi_1\, h_4+\phi_2\, h_5+\phi_3\, h_6)\right]
\exp \Bigl[-\zeta n_{13}^s \Bigr]\exp\Bigl[-\frac{i}{2}\,r \,\gamma^s_1 \Bigr]\,.
\end{eqnarray} 
By solving the equations in (\ref{PJ-CYBE})\,, the deformed current 
$J^{\hat{\ga}_1,\hat{\ga}_2,\hat{\ga}_3}_\alpha$ is determined as
\begin{eqnarray}
J^{\hat{\ga}_1,\hat{\ga}_2,\hat{\ga}_3}_\pm&=&-i\partial_\pm r\,\frac{\ga_1^s}{2}
-\partial_\pm \zeta\left[i\sin r\,\frac{\ga_3^s}{2}+\cos r\,n_{13}^s\right]\nonumber \\
&&-G(\hat{\gamma}_i)\Bigl[\partial_\pm\phi_1\pm(\hat{\gamma}_3\sin^2 r\sin^2\zeta\partial_\pm\phi_2
-\hat{\gamma}_2\cos^2 r\partial_\pm\phi_3)\nonumber \\
&&\qquad\qquad\qquad\qquad\qquad\qquad+\hat{\gamma}_1\sin^2 r\cos^2 r\sin^2\zeta
\sum_{i=1}^3\hat{\gamma}_i\partial_\pm\phi_i\Bigr]\nonumber \\
&&\qquad\qquad\qquad\qquad\times\left[\cos\zeta(i\sin r\,\frac{\ga_2^s}{2}+\cos r\,n_{12}^s)
+\sin \zeta\,n_{23}^s\right]
\nonumber \\
&&-G(\hat{\gamma}_i)\Bigl[\partial_\pm\phi_2\pm(\hat{\gamma}_1\cos^2 r\partial_\pm\phi_3
-\hat{\gamma}_3\sin^2 r\cos^2\zeta\partial_\pm\phi_1)\nonumber \\
&&\qquad\qquad\qquad\qquad\qquad\qquad+\hat{\gamma}_2\sin^2 r\cos^2 r\cos^2\zeta
\sum_{i=1}^3\hat{\gamma}_i\partial_\pm\phi_i\Bigr]\nonumber \\
&&\qquad\qquad\qquad\qquad\times\left[\sin\zeta(i\,\sin r\,\frac{1}{2}\ga_4^s+\cos r\,n_{14}^s)
+\cos\zeta\,n_{34}^s\right]\nonumber \\
&&+G(\hat{\gamma}_i)\Bigl[\partial_\pm\phi_3\pm(\hat{\gamma}_2\sin^2 r\cos^2\zeta\partial_\pm\phi_1
-\hat{\gamma}_1\sin^2 r\sin^2\zeta\partial_\pm\phi_2)\nonumber \\
&&\qquad\qquad\qquad\qquad\qquad\qquad+\hat{\gamma}_3\sin^4 r\sin^2\zeta\cos^2\zeta
\sum_{i=1}^3\hat{\gamma}_i\partial_\pm\phi_i\Bigr]\nonumber \\
&&\qquad\qquad\qquad\qquad\times\left[i\cos r\,\frac{\ga_5^s}{2}-\sin r\,n_{15}^s\right]\,.
\label{3p-current-LM}
\end{eqnarray}
Here the parameters $\hat{\gamma}_i$ are defined as 
\begin{eqnarray}
\hat{\gamma}_i \equiv 8\,\eta\mu_i\,, 
\end{eqnarray}
and the scalar function $G(\hat{\gamma}_i)$ is 
\begin{eqnarray}
G^{-1}(\hat{\gamma}_i)& \equiv&1+\sin^2 r(\hat{\gamma}_1^2\cos^2 r\sin^2\zeta+\hat{\gamma}_2^2\cos^2 r\cos^2\zeta
+\hat{\gamma}_3^2\sin^2 r\sin^2\zeta\cos^2\zeta)\,. 
\end{eqnarray}
This deformed current (\ref{3p-current-LM}) will play an important role in the following analysis. 

\medskip 

Substituting the deformed current (\ref{3p-current-LM}) into the Lagrangian (\ref{L}) 
leads to the background 
\begin{eqnarray}
ds^2 &=& \sum_{i=1}^{3} \left( {d\rho_i}^2+G(\hat{\gamma}_i) {\rho_i}^2 {d\phi_i}^2 \right)
+G(\hat{\gamma}_i){\rho_1}^2 {\rho_2}^2 {\rho_3}^2 \left( \sum_{i=1}^{3} \hat{\gamma}_i d{\phi_i} \right)^2\,,\nonumber\\
B_2&=&G(\hat{\gamma}_i)\,(\hat{\gamma}_3{\rho_1}^2{\rho_2}^2 d\phi_1\wedge d\phi_2
+\hat{\gamma}_1 {\rho_2}^2{\rho_3}^2 d\phi_2\wedge d\phi_3+\hat{\gamma}_2{\rho_3}^2{\rho_1}^2 d\phi_3\wedge d\phi_1)\,.
\label{3p-Lunin-Maldacena}
\end{eqnarray}
Here new coordinates $\rho_i~(i=1,2,3)$ are defined as
\begin{eqnarray}
\rho_1\equiv \sin r\cos  \zeta\,,\qquad \rho_2 \equiv \sin r\sin\zeta\,,\qquad\rho_3 \equiv\cos r\,. 
\end{eqnarray}
The metric and NS-NS two-form in (\ref{3p-Lunin-Maldacena}) agree
with 3-parameter $\gamma$-deformations of S$^5$ \cite{Frolov}. 

\medskip 

A particular one-parameter case with 
\begin{eqnarray}
\hat{\gamma}_1=\hat{\gamma}_2=\hat{\gamma}_3 \equiv \hat{\gamma}  
\end{eqnarray}
corresponds to the Lunin-Maldacena solution \cite{LM} described by 
\begin{eqnarray}
ds^2 &=& \sum_{i=1}^{3} \left( {d\rho_i}^2+G {\rho_i}^2 {d\phi_i}^2 \right)
+G \hat{\gamma}^2{\rho_1}^2 {\rho_2}^2 {\rho_3}^2 \left( \sum_{i=1}^{3}d{\phi_i}\right)^2\,,\nonumber\\
B_2&=&G\,\hat{\gamma}({\rho_1}^2{\rho_2}^2 d\phi_1\wedge d\phi_2
+{\rho_2}^2{\rho_3}^2 d\phi_2\wedge d\phi_3+{\rho_3}^2{\rho_1}^2 d\phi_3\wedge d\phi_1)\,, 
\label{Lunin-Maldacena}
\end{eqnarray}
where the scalar function $G$ is defined as 
\begin{eqnarray}
G^{-1} \equiv 1+\frac{\hat{\gamma}^2}{4}(\sin^22r+\sin^4r\sin^22\zeta)\,.
\end{eqnarray}
This background is a holographic dual of the $\beta$-deformation of 
the $\mathcal{N}=4$ super Yang-Mills theory \cite{LS}.

\subsubsection*{Lax pair}

The next task is to evaluate the Lax pair in (\ref{lax0}) with the classical $r$-matrix (\ref{abelian})\,. 
The components $\mL^{\hat{\ga}_1,\hat{\ga}_2,\hat{\ga}_3}_\pm$ are given by 
\begin{eqnarray}
\mL^{\hat{\ga}_1,\hat{\ga}_2,\hat{\ga}_3}_\pm&=&-i\frac{\la^{\pm1}}{2}\partial_\pm r\,\ga_1^s
-\partial_\pm \zeta\left[i\sin r\,\frac{\la^{\pm1}}{2}\ga_3^s+\cos r\,n_{13}^s\right]\nonumber \\
&&-G(\hat{\ga}_i)\Bigl[\partial_\pm\phi_1\pm(\hat{\ga}_3\sin^2 r\sin^2\zeta\partial_\pm
\phi_2-\hat{\ga}_2\cos^2 r\partial_\pm\phi_3)\nonumber \\
&&\qquad\qquad\qquad\qquad\qquad\qquad+\hat{\ga}_1\sin^2 r\cos^2 r\sin^2\zeta
\sum_{i=1}^3\hat{\ga}_i\partial_\pm\phi_i\Bigr]\nonumber \\
&&\qquad\qquad\qquad\qquad\times\left[\cos\zeta(i\sin r\,\frac{\la^{\pm1}}{2}\ga_2^s+\cos r\,n_{12}^s)
+\sin \zeta\,n_{23}^s\right]\nonumber \\
&&-G(\hat{\ga}_i)\Bigl[\partial_\pm\phi_2\pm(\hat{\ga}_1\cos^2 r
\partial_\pm\phi_3-\hat{\ga}_3\sin^2 r\cos^2\zeta\partial_\pm\phi_1)
\nonumber \\
&&\qquad\qquad\qquad\qquad\qquad\qquad+\hat{\ga}_2\sin^2 r\cos^2 r\cos^2\zeta
\sum_{i=1}^3\hat{\ga}_i\partial_\pm\phi_i\Bigr]\nonumber \\
&&\qquad\qquad\qquad\qquad\times\left[\sin\zeta(i\,\sin r\,\frac{\la^{\pm1}}{2}\ga_4^s+\cos r\,n_{14}^s)
+\cos\zeta\,n_{34}^s\right]\nonumber \\
&&+G(\hat{\ga}_i)\Bigl[\partial_\pm\phi_3\pm(\hat{\ga}_2\sin^2 r\cos^2\zeta\partial_\pm\phi_1
-\hat{\ga}_1\sin^2 r\sin^2\zeta\partial_\pm\phi_2)\nonumber \\
&&\qquad\qquad\qquad\qquad\qquad\qquad+\hat{\ga}_3\sin^4 r\sin^2\zeta\cos^2\zeta
\sum_{i=1}^3\hat{\ga}_i\partial_\pm\phi_i\Bigr]\nonumber \\
&&\qquad\qquad\qquad\qquad\times\left[i\cos r\,\frac{\la^{\pm1}}{2}\ga_5^s-\sin r\,n_{15}^s\right]\,.
\label{3p-lax-gamma}
\end{eqnarray} 

Note here that, in the undeformed limit 
$\hat{\ga}_i\to\,0$, $\mL^{\hat{\ga}_1,\hat{\ga}_2,\hat{\ga}_3}_\pm$ become
\begin{eqnarray}
\mL_\pm^{S}&=&-i\,\partial_\pm r \frac{\la^{\pm1}}{2}\ga_1^s-\partial_\pm\zeta\,
\left[i\sin r\,\frac{\la^{\pm1}}{2}\ga_3^s+\cos r\,n_{13}^s\right] 
\nonumber \\
&&-\partial_\pm\phi_1\left[\cos\zeta(i\sin r\,\frac{\la^{\pm1}}{2}\ga_2^s+\cos r\,n_{12}^s)
+\sin \zeta\,n_{23}^s\right]\nonumber \\
&&-\,\partial_\pm\phi_2\left[\sin\zeta(i\,\sin r\,\frac{\la^{\pm1}}{2}\ga_4^s+\cos r\,n_{14}^s)
+\cos\zeta\,n_{34}^s\right]\nonumber \\
&&+\partial_{\pm}\phi_3\,\left[i\cos r\,\frac{\la^{\pm1}}{2}\ga_5^s-\sin r\,n_{15}^s\right]\,. 
\label{ud-LM}
\end{eqnarray}
This is just a Lax pair for the undeformed $\rm{S}^5$.

\subsubsection*{Another derivation of Lax pair}

Then, let us consider a simple derivation of the Lax pair (\ref{3p-lax-gamma})\,, as in the previous section. 
The undeformed current is given by 
\begin{eqnarray}
A_\pm&=&-i\partial_\pm r\,\frac{\ga_1^s}{2}
-\partial_\pm \zeta\left[i\sin r\,\frac{\ga_3^s}{2}+\cos r\,n_{13}^s\right]\nonumber \\
&&-\partial_\pm\phi_1\left[\cos\zeta(i\sin r\,\frac{\ga_2^s}{2}+\cos r\,n_{12}^s)
+\sin \zeta\,n_{23}^s\right]\nonumber \\
&&-\partial_\pm\phi_2\left[\sin\zeta(i\,\sin r\,\frac{\ga_4^s}{2}+\cos r\,n_{14}^s)
+\cos\zeta\,n_{34}^s\right]\nonumber \\
&&+\partial_\pm\phi_3\left[i\cos r\,\frac{\ga_5^s}{2}-\sin r\,n_{15}^s\right]\,.
\label{udc-LM}
\end{eqnarray}
By comparing the deformed current (\ref{3p-current-LM}) with the undeformed one (\ref{udc-LM})\,, 
we can identify the following replacement rules: 
\begin{eqnarray}
\partial_\pm\phi_1 &~\longrightarrow~&
G(\hat{\ga}_i)\Bigl[\partial_\pm\phi_1\pm(\hat{\ga}_3\sin^2 r\sin^2\zeta\partial_\pm\phi_2
-\hat{\ga}_2\cos^2 r\partial_\pm\phi_3)\nonumber \\
&&\qquad\qquad\qquad\qquad\qquad\qquad+\hat{\ga}_1\sin^2 r\cos^2 r\sin^2\zeta
\sum_{i=1}^3\hat{\ga}_i\partial_\pm\phi_i\Bigr]\,,\nonumber\\
\partial_\pm\phi_2 &~\longrightarrow~& 
G(\hat{\ga}_i)\Bigl[\partial_\pm\phi_2\pm(\hat{\ga}_1\cos^2 r\partial_\pm\phi_3
-\hat{\ga}_3\sin^2 r\cos^2\zeta\partial_\pm\phi_1)\,,\nonumber \\
&&\qquad\qquad\qquad\qquad\qquad\qquad+\hat{\ga}_2\sin^2 r\cos^2 r\cos^2\zeta
\sum_{i=1}^3\hat{\ga}_i\partial_\pm\phi_i\Bigr]\,,\nonumber\\
\partial_\pm\phi_3 
&~\longrightarrow~& G(\hat{\ga}_i)\Bigl[\partial_\pm\phi_3
\pm(\hat{\ga}_2\sin^2 r\cos^2\zeta\partial_\pm\phi_1
-\hat{\ga}_1\sin^2 r\sin^2\zeta\partial_\pm\phi_2)\nonumber \\
&&\qquad\qquad\qquad\qquad\qquad\qquad+\hat{\ga}_3\sin^4 r\sin^2\zeta\cos^2\zeta
\sum_{i=1}^3\hat{\ga}_i\partial_\pm\phi_i\Bigr]\,.
\label{3p-TsTvari}
\end{eqnarray} 
Due to the replacement rules (\ref{3p-TsTvari})\,, the deformed Lax pair (\ref{3p-lax-gamma}) 
can be reconstructed from the undeformed one (\ref{ud-LM})\,.
In fact, the replacement rules (\ref{3p-TsTvari}) are identical to a TsT-transformation 
\cite{Frolov}\footnote{In our argument, the rules are identified on the off-shell level, 
but the one in \cite{Frolov} is done on the on-shell.}, and hence the Lax pair (\ref{3p-lax-gamma}) 
is equivalent to the one in \cite{Frolov}\footnote{
To see this equivalence (up to small differences of convention), 
we have to perform a gauge transformation
\[
h=\exp[-i\zeta n_{02}]\exp\Bigl[\frac{i}{2}r\ga_2\Bigr]\exp\Bigl[\frac{\pi}{2}(n_{12}+i n_{03})\Bigr]
\]
and a M\"obius transformation for a spectral parameter $\la\to\frac{\la+1}{\la-1}$\,.}. 
This result further confirms the correspondence between a Yang-Baxter deformation with (\ref{abelian}) 
and a TsT transformation \cite{MY1}.  

\medskip 

Finally, it is worth mentioning the reinterpretation of the deformation 
as a twisted periodic boundary condition. This fact was originally shown in \cite{Frolov}. 
The twisted periodic boundary condition with the undeformed 
AdS$_5\times$S$^5$ is given by 
\begin{eqnarray}
\tilde{\phi}_1(\sigma=2\pi)&=&\tilde{\phi}_1(\sigma=0)
+\gamma_3J_2 - \gamma_2J_3+2\pi n_1\,, \nonumber\\
\tilde{\phi}_2(\sigma=2\pi)&=&\tilde{\phi}_2(\sigma=0)
+\gamma_1J_3 - \gamma_3J_1+2\pi n_2\,, \nonumber\\
\tilde{\phi}_3(\sigma=2\pi)&=&\tilde{\phi}_3(\sigma=0)
+\gamma_2J_1 - \gamma_1 J_2+2\pi n_3\,,
\end{eqnarray}
with $\gamma_i \equiv \hat{\gamma}/\sqrt{\lambda_{\rm c}}$\,. 
Here $J_i$ are Noether charges for rotation invariance 
in the $\phi_i$ directions. 
Integers $n_i$ are winding numbers along the $\phi_i$ directions.

\section{Lax pairs for Schr\"odinger spacetimes}

Let us consider a classical $r$-matrix which deforms both AdS$_5$ and S$^5$\,. 
Such an $r$-matrix contains generators of both $\mathfrak{su}(2,2)$ and $\mathfrak{su}(4)$\,.  
A simple example is the following \cite{Sch-YB}:
\begin{eqnarray}
r=\frac{i}{4\sqrt{2}}\,(p_0-p_3)\wedge(h_4+h_5+h_6)\,. \label{r-Sch}
\end{eqnarray}
For convention of the generators, see Appendix A. 
This $r$-matrix (\ref{r-Sch}) is associated with Schr\"odinger spacetimes realized in type IIB supergravity 
\cite{10DSch}, as shown in \cite{Sch-YB}. 

\subsubsection*{The deformed metric and NS-NS two-form}

The bosonic group elements of $SU(2,2)$ and $SU(4)$ are parameterized as
\begin{eqnarray}
g_a(\tau,\sigma)&=&\exp\left[x^0p_0+x^1p_1+x^2p_2+x^3p_3\right]
\exp\left[\ga^a_5\,\frac{1}{2}\,\log z\right] ~~\in~ SU(2,2)\,,\nonumber\\
g_s(\tau,\sigma)&=&\exp\left[\frac{i}{2}(\psi_1h_4+\psi_2h_5+\psi_3h_6)\right]
\exp\Bigl[-\zeta n^s_{13}\Bigr]\exp\left[-\frac{i}{2}r\,\gamma^s_1\right] ~~\in~ SU(4)\,. 
\label{S5}
\end{eqnarray}
The deformed current $J_\pm$ can be expanded in terms of the generators of $\alg{su}(2,2)\oplus\alg{su}(4)$\,. 
Then, by solving the equations in (\ref{PJ-CYBE})\,, $J_{\pm}$ is determined as  
\begin{eqnarray}
\label{current-Sch}
J^a_{\pm}&=&\frac{1}{z}\partial_{\pm}x^1p_1
+\frac{1}{z}\partial_{\pm}x^2p_2
+\frac{1}{2z}\partial_{\pm}z\,\gamma^a_5\nonumber \\
&&+\frac{1}{\sqrt{2}\,z}\partial_{\pm}x^+\left(p_0+p_3\right) \nonumber\\
&&+\frac{1}{\sqrt{2}\,z}\left[\partial_{\pm}x^- 
\pm\eta\partial_{\pm}\chi\pm\frac{\eta}{2}\sin^2\mu(\partial_\pm\psi
+\cos\theta\partial_\pm\phi)+\frac{\eta^2}{z^2}\partial_{\pm}x^+\right](p_0-p_3)\,,\nonumber\\
J^s_{\pm}&=&-\frac{i}{2}\partial_{\pm}\mu\,\gamma^s_1
-\frac{1}{2}\partial_{\pm}\theta\left[\frac{i}{2}\sin\mu\,\gamma^s_3+\cos\mu\, 
n^s_{13}\right]\nonumber\\
&&-\left[\partial_\pm\chi\pm\frac{\eta\,\partial_{\pm}x^+}{z^2}
\right]\biggl[\sin\frac{\theta}{2}\left(\frac{i}{2}\sin\mu\,\ga^s_4+\cos\mu \,
n^s_{14}+n^s_{23}\right)\nonumber \\
&&\qquad\qquad+\cos\frac{\theta}{2}\left(\frac{i}{2}\sin\mu\,\ga^s_2
+\cos\mu \,n^s_{12}+n^s_{34}\right)
-\frac{i}{2}\cos\mu\,\ga_5^s+\sin\mu\,n^s_{15}\biggr]\nonumber \\
&&+\frac{1}{2}\partial_\pm\phi\biggl[\sin\frac{\theta}{2}\left(\frac{i}{2}\sin\mu\,\ga^s_4+\cos\mu \,
n^s_{14}-n^s_{23}\right)\nonumber \\
&&\qquad\qquad\qquad
-\cos\frac{\theta}{2}\left(\frac{i}{2}\sin\mu\,\ga^s_2
+\cos\mu \,n^s_{12}-n^s_{34}\right)\biggr]\nonumber \\
&&-\frac{1}{2}\partial_\pm\psi\biggl[\sin\frac{\theta}{2}\left(\frac{i}{2}\sin\mu\,\ga^s_4+\cos\mu \,
n^s_{14}+n^s_{23}\right)\nonumber \\
&&\qquad\qquad\qquad+\cos\frac{\theta}{2}\left(\frac{i}{2}\sin\mu\,\ga^s_2
+\cos\mu \,n^s_{12}+n^s_{34}\right)\biggr]\,. \label{dc-Sch}
\end{eqnarray}
Here we have performed a coordinate transformation, 
\begin{eqnarray}
&x^\pm&= \frac{x^0\pm x^3}{\sqrt{2}}\,,\nonumber \\
&r&= \mu\,, \quad \ze= \tfrac{1}{2}\theta\,, \quad 
\psi_1 = \chi+\tfrac{1}{2}(\psi+\phi)\,, \quad \psi_2= \chi+\tfrac{1}{2}(\psi-\phi)\,, 
\quad \psi_3=\chi\,.  \nonumber 
\end{eqnarray}
With the deformed current (\ref{dc-Sch})\,, the resulting background is given by 
\begin{eqnarray}
ds^2&=&\frac{-2dx^+dx^-+(dx^1)^2+(dx^2)^2+dz^2}{z^2}-\eta^2\frac{(dx^+)^2}{z^4}
+ds_{\txt{S}^5}^2\,,\nonumber \\
B_2&=&\frac{\eta}{z^2}\,dx^+\wedge (d\chi+\omega)\,. 
\end{eqnarray}
Here the S$^5$ metric is written as an $S^1$-fibration over $\mathbb{C}\text{P}^2$\,, 
\begin{align} 
ds^2_{\rm S^5}&=(d\chi+\omega)^2 +ds^2_{\rm \mathbb{C}P^2}\,, \nln 
ds^2_{\rm \mathbb{C}P^2}&= d\mu^2+\sin^2\mu\,
\bigl(\Sigma_1^2+\Sigma_2^2+\cos^2\mu\,\Sigma_3^2\bigr)\,. 
\label{S1overCP2}
\end{align}
Now $\chi$ is the fiber coordinate and $\omega$ is 
a one-form potential of the K\"ahler form on $\mathbb{C}$P$^2$\,. 
The symbols $\Sigma_i ~(i=1,2,3)$ and $\omega$ are defined as  
\begin{align}
\Sigma_1&\equiv \tfrac{1}{2}(\cos\psi\, d\theta +\sin\psi\sin\theta\, d\phi)\,, \nln 
\Sigma_2&\equiv \tfrac{1}{2}(\sin\psi\, d\theta -\cos\psi\sin\theta\, d\phi)\,, \nln 
\Sigma_3&\equiv \tfrac{1}{2}(d\psi +\cos\theta\, d\phi)\,, 
\qquad 
\omega \equiv \sin^2\mu\, \Sigma_3\,. 
\end{align}
It is remarkable that only the AdS$_5$ metric is deformed while the S$^5$ part is not, 
in spite of the expression of the classical $r$-matrix (\ref{r-Sch})\,. 
On the other hand, the NS-NS two-form carries two indices, one of which is from AdS$_5$ and the other is S$^5$\,.

\subsubsection*{Lax pair}

In a similar way, one can evaluate the associated Lax pair. The resulting expression is a bit messy 
but given by  
\begin{eqnarray}
\mL^{\rm{Sch}}_{\pm}&=&\frac{1}{z}\partial_{\pm}x^1\left[\frac{\lambda^{\pm1}}{2}\gamma^a_1-n^a_{15}\right]
+\frac{1}{z}\partial_{\pm}x^2\left[\frac{\lambda^{\pm1}}{2}\gamma^a_2-n^a_{25}\right]
+\frac{\lambda^{\pm1}}{2z}\partial_{\pm}z\,\gamma^a_5\nonumber \\
&&+\frac{1}{\sqrt{2}\,z}\partial_{\pm}x^+\left[\frac{\lambda^{\pm1}}{2}\gamma^a_0
+\frac{\lambda^{\pm1}}{2}\gamma^a_3-n^a_{05} -n^a_{35}\right]\nonumber\\
&&+\frac{1}{\sqrt{2}\,z}\left[\partial_{\pm}x^- 
\pm\eta\partial_{\pm}\chi\pm\frac{\eta}{2}\sin^2\mu(\partial_\pm\psi+\cos\theta\partial_\pm\phi)
+\frac{\eta^2}{z^2}\partial_{\pm}x^+\right]\nonumber \\
&&\qquad\qquad\quad\times\left[\frac{\lambda^{\pm1}}{2}\gamma^a_0-\frac{\lambda^{\pm1}}{2}
\gamma^a_3-n^a_{05} +n^a_{35}\right]\nonumber\\
&&-\frac{i\lambda^{\pm1}}{2}\partial_{\pm}\mu\,\gamma^s_1
-\frac{1}{2}\partial_{\pm}\theta\left[\frac{i\lambda^{\pm1}}{2}\sin\mu\,\gamma^s_3+\cos\mu\, 
n^s_{13}\right]\nonumber\\
&&-\left[\partial_\pm\chi\pm\frac{\eta\,\partial_{\pm}x^+}{z^2}
\right]\,\biggl[\sin\frac{\theta}{2}\left(\frac{i\lambda^{\pm1}}{2}\sin\mu\,\ga^s_4+\cos\mu \,
n^s_{14}+n^s_{23}\right)\nonumber \\
&&\qquad\qquad+\cos\frac{\theta}{2}\left(\frac{i\lambda^{\pm1}}{2}\sin\mu\,\ga^s_2
+\cos\mu \,n^s_{12}+n^s_{34}\right)
-\frac{i\lambda^{\pm1}}{2}\cos\mu\,\ga_5^s+\sin\mu\,n^s_{15}\biggr]\nonumber \\
&&+\frac{1}{2}\partial_\pm\phi\biggl[\sin\frac{\theta}{2}
\left(\frac{i\lambda^{\pm1}}{2}\sin\mu\,\ga^s_4+\cos\mu \,
n^s_{14}-n^s_{23}\right)\nonumber \\
&&\qquad\qquad\qquad
-\cos\frac{\theta}{2}\left(\frac{i\lambda^{\pm1}}{2}\sin\mu\,\ga^s_2
+\cos\mu \,n^s_{12}-n^s_{34}\right)\biggr]\nonumber \\
&&-\frac{1}{2}\partial_\pm\psi\biggl[\sin\frac{\theta}{2}
\left(\frac{i\lambda^{\pm1}}{2}\sin\mu\,\ga^s_4+\cos\mu \,
n^s_{14}+n^s_{23}\right)\nonumber \\
&&\qquad\qquad\qquad+\cos\frac{\theta}{2}\left(\frac{i\lambda^{\pm1}}{2}\sin\mu\,\ga^s_2
+\cos\mu \,n^s_{12}+n^s_{34}\right)\biggr]\,. 
\label{lax-Sch}
\end{eqnarray} 
It would be helpful to check the undeformed limit.  
As $\eta\to0$\,, the above Lax pair $\mL^{\text{Sch}}_\pm$ is reduced to the following: 
\begin{eqnarray}
\mL_{\pm}&=&\frac{1}{z}\partial_{\pm}x^1\left[\frac{\lambda^{\pm1}}{2}\gamma^a_1-n^a_{15}\right]
+\frac{1}{z}\partial_{\pm}x^2\left[\frac{\lambda^{\pm1}}{2}\gamma^a_2-n^a_{25}\right]
+\frac{\lambda^{\pm1}}{2z}\partial_{\pm}z\,\gamma^a_5\nonumber \\
&&+\frac{1}{\sqrt{2}\,z}\partial_{\pm}x^+\left[\frac{\lambda^{\pm1}}{2}\gamma^a_0
+\frac{\lambda^{\pm1}}{2}\gamma^a_3-n^a_{05} -n^a_{35}\right]\nonumber\\
&&+\frac{1}{\sqrt{2}\,z}\partial_{\pm}x^-
\left[\frac{\lambda^{\pm1}}{2}\gamma^a_0-\frac{\lambda^{\pm1}}{2}
\gamma^a_3-n^a_{05} +n^a_{35}\right]\nonumber\\
&&-\frac{i\lambda^{\pm1}}{2}\partial_{\pm}\mu\,\gamma^s_1
-\frac{1}{2}\partial_{\pm}\theta\left[\frac{i\lambda^{\pm1}}{2}\sin\mu\,\gamma^s_3+\cos\mu\, 
n^s_{13}\right]\nonumber\\
&&-\partial_\pm\chi
\biggl[\sin\frac{\theta}{2}\left(\frac{i\lambda^{\pm1}}{2}\sin\mu\,\ga^s_4+\cos\mu \,
n^s_{14}+n^s_{23}\right)\nonumber \\
&&\qquad\qquad+\cos\frac{\theta}{2}\left(\frac{i\lambda^{\pm1}}{2}\sin\mu\,\ga^s_2
+\cos\mu \,n^s_{12}+n^s_{34}\right)
-\frac{i\lambda^{\pm1}}{2}\cos\mu\,\ga_5^s+\sin\mu\,n^s_{15}\biggr]\nonumber \\
&&+\frac{1}{2}\partial_\pm\phi\biggl[\sin\frac{\theta}{2}
\left(\frac{i\lambda^{\pm1}}{2}\sin\mu\,\ga^s_4+\cos\mu \,
n^s_{14}-n^s_{23}\right)\nonumber \\
&&\qquad\qquad\qquad
-\cos\frac{\theta}{2}\left(\frac{i\lambda^{\pm1}}{2}\sin\mu\,\ga^s_2
+\cos\mu \,n^s_{12}-n^s_{34}\right)\biggr]\nonumber \\
&&-\frac{1}{2}\partial_\pm\psi\biggl[\sin\frac{\theta}{2}
\left(\frac{i\lambda^{\pm1}}{2}\sin\mu\,\ga^s_4+\cos\mu \,
n^s_{14}+n^s_{23}\right)\nonumber \\
&&\qquad\qquad\qquad+\cos\frac{\theta}{2}\left(\frac{i\lambda^{\pm1}}{2}\sin\mu\,\ga^s_2
+\cos\mu \,n^s_{12}+n^s_{34}\right)\biggr]\,. 
\label{ud-lax-Sch}
\end{eqnarray}

\subsubsection*{Another derivation of Lax pair}

Let us consider another derivation of the Lax pair again. 
One can see the replacement rules by comparing the deformed current with the undeformed one, as before. 

\medskip 

The undeformed current is decomposed into the AdS$_5$ and S$^5$ components like 
\begin{eqnarray}
\label{udc-Sch}
A_{\pm}&=&A^a_{\pm}+A^s_{\pm}\,, \label{uc}
\end{eqnarray}
where $A^a_{\pm}$ and $A^s_{\pm}$ are given by 
\begin{eqnarray}
A^a_{\pm}&=&\frac{1}{z}\partial_{\pm}x^1p_1
+\frac{1}{z}\partial_{\pm}x^2p_2
+\frac{1}{2z}\partial_{\pm}z\,\gamma^a_5+\frac{1}{\sqrt{2}\,z}\partial_{\pm}x^+\left(p_0+p_3\right)
+\frac{1}{\sqrt{2}\,z}\partial_{\pm}x^- (p_0-p_3)\,, \nonumber \\
A^s_{\pm}&=&-\frac{i}{2}\partial_{\pm}\mu\,\gamma^s_1
-\frac{1}{2}\partial_{\pm}\theta\left[\frac{i}{2}\sin\mu\,\gamma^s_3+\cos\mu\, n^s_{13}\right]\nonumber\\
&&-\partial_\pm\chi\biggl[\sin\frac{\theta}{2}\left(\frac{i}{2}\sin\mu\,\ga^s_4+\cos\mu \,
n^s_{14}+n^s_{23}\right) \nonumber \\
&&\qquad\qquad+\cos\frac{\theta}{2}\left(\frac{i}{2}\sin\mu\,\ga^s_2
+\cos\mu \,n^s_{12}+n^s_{34}\right)
-\frac{i}{2}\cos\mu\,\ga_5^s+\sin\mu\,n^s_{15}\biggr]\nonumber \\
&&+\frac{1}{2}\partial_\pm\phi\biggl[\sin\frac{\theta}{2}\left(\frac{i}{2}\sin\mu\,\ga^s_4+\cos\mu \,
n^s_{14}-n^s_{23}\right)\nonumber \\
&&\qquad\qquad\qquad
-\cos\frac{\theta}{2}\left(\frac{i}{2}\sin\mu\,\ga^s_2+\cos\mu \,n^s_{12}-n^s_{34}\right)\biggr]\nonumber \\
&&-\frac{1}{2}\partial_\pm\psi\biggl[\sin\frac{\theta}{2}\left(\frac{i}{2}\sin\mu\,\ga^s_4+\cos\mu \,
n^s_{14}+n^s_{23}\right)\nonumber \\
&&\qquad\qquad\qquad+\cos\frac{\theta}{2}\left(\frac{i}{2}\sin\mu\,\ga^s_2
+\cos\mu \,n^s_{12}+n^s_{34}\right)\biggr]\,. \label{uc2}
\end{eqnarray}
Thus, by comparing the deformed current (\ref{dc-Sch}) with the undeformed one (\ref{uc2}), 
one can see the following replacement rules:
\begin{eqnarray}
\frac{1}{z^2}\,\partial_\pm x^- & ~~\longrightarrow~~& \frac{1}{z^2}\,\partial_\pm x^-\pm \frac{\eta}{z^2}
\Bigl[\partial_\pm\chi+\frac{1}{2}\sin^2\mu\,(\partial_\pm\psi+\cos\theta\partial_\pm\phi)\Bigr]
+\frac{\eta^2}{z^4}\,\partial_\pm x^+\,, \nonumber\\
\partial_\pm \chi &~~\longrightarrow~~& \partial_\pm \chi\pm\frac{\eta}{z^2}\,\partial_\pm x^+\,. \label{rule-3}
\end{eqnarray}
Then the Lax pair (\ref{lax-Sch}) can be reproduced by applying the replacement rules (\ref{rule-3})
to the undeformed Lax pair (\ref{ud-lax-Sch})\,. 

\medskip 

Finally, let us mention about the reinterpretation of the deformation as a twisted periodic boundary condition. 
Similarly, one can see that the following boundary condition 
\begin{eqnarray}
\tilde{x}^-(\sigma=2\pi)&=&\tilde{x}^-(\sigma=0)+\frac{\eta}{\sqrt{\lambda_{\rm c}}}{J_\chi}+2\pi n_\chi\,, \nonumber \\
\tilde{\chi}(\sigma=2\pi)&=&\tilde{\chi}(\sigma=0)-\frac{\eta}{\sqrt{\lambda_{\rm c}}}{P_-}
\end{eqnarray}
with the undeformed AdS$_5\times$S$^5$ is equivalent to the deformed geometry with the usual periodic 
boundary condition. Here $P_-$ and $J_\chi$ are Noether charges for translation and rotation invariance 
for the $x^-$ and $\chi$ directions, respectively. 
An integer $n_\chi$ is a winding number for the $\chi$ direction. 
It may be interesting to consider a relation between the above argument 
and the symmetric two-form studied in \cite{SYY}.

\section{Lax pairs for abelian twists of the global AdS$_5$}

Finally, we will consider abelian twists of the global AdS$_5$ as Yang-Baxter deformations.
The twists are associated with the classical $r$-matrix,
\begin{eqnarray}
r=-\frac{i}{2}n^a_{12}\wedge n^a_{03}\,. 
\end{eqnarray}
This is composed of two Cartan generators of $\mathfrak{su}(2,2)$ (For our convention, see Appendix A). 
This $r$-matrix deforms only the AdS$_5$ part, hence we will omit the S$^5$ part hereafter.

\subsubsection*{The deformed metric and NS-NS two-form}

We will work with the following parameterization of a group element of $SU(2,2)$:
\begin{eqnarray}
g_a(\tau,\sigma)=\exp\left[\frac{i}{2}(\phi_1\,h_1+\phi_2\,h_2+\tau\,h_3)\right]
\exp\left[-\theta\,n_{13}^a\right]\exp\left[-\rho\,\frac{\ga_1^a}{2}\right]
\quad
\in SU(2,2)\,.
\end{eqnarray}
The deformed current $J_\pm$ is expanded in terms of the basis of $\alg{su}(2,2)$\,. 
Then, by solving the equations in (\ref{PJ-CYBE})\,,  $J_{\pm}$ can be determined as 
\begin{eqnarray}
\label{current-DHH}
J_{\pm}&=&-\partial_\pm\rho\frac{1}{2}\gamma^a_1
-\partial_\pm\theta\left[\frac{1}{2}\sinh\rho\,\gamma^a_3+\cosh\rho\,n^a_{13}\right]\nonumber\\
&&
+i\partial_\pm\tau \left[\frac{1}{2}\cosh\rho\,\gamma^a_5
+\sinh\rho\,n^a_{15}\right]\nonumber\\
&&-\hat{G}\left(\partial_\pm\phi_1\mp\eta\sin^2\theta\sinh^2\rho\,\partial_\pm\phi_2\right)\nonumber \\
&&\qquad\qquad\qquad\times\left[\cos\theta\left(\frac{1}{2}\sinh\rho\,
\gamma^a_2+\cosh\rho\,n_{12}^a\right)+\sin\theta\,n_{23}^a\right]
\nonumber\\
&&+i\hat{G}\left(\partial_\pm\phi_2\pm\eta\cos^2\theta\sinh^2\rho\,\partial_\pm\phi_1\right) \nonumber \\
&&\qquad\qquad\qquad\times\left[\sin\theta\left(\frac{1}{2}\sinh\rho\,\ga_0^a-\cosh\rho\,n_{01}^a\right)
-\cos\theta \,n^a_{03}\right]\,, \label{dc-ga}
\end{eqnarray}
where $\hat{G}$ is a scalar function defined as
\begin{eqnarray}
\hat{G}^{-1} \equiv 1+\eta^2\sin^2\theta\cos^2\theta\sinh^2\rho\,.
\end{eqnarray}
By using the current (\ref{dc-ga})\,, 
the deformed metric and NS-NS two-form are obtained as  
\begin{eqnarray}
ds^2&=&-\cosh^2\rho\,d\tau^2+d\rho^2+\sinh ^2\rho\,
\Bigl(d\theta^2+\frac{\cos^2\theta\,d\phi_1^2 
+\sin^2\theta\,d\phi_2^2}{1+\eta^2 \sin ^2\theta\cos^2\theta\sinh^4\rho}\Bigr)\,,
\nonumber\\
B_2&=&-\frac{\eta\sin^2\theta\cos^2\theta\sinh^4\rho}{1+\eta^2 \sin^2\theta\cos ^2\theta\sinh^4\rho}
d\phi_1\wedge d\phi_2\,.
\label{abe-twst}
\end{eqnarray}
This result precisely agrees with abelian twists of the global AdS$_5$ \cite{MS}.

\subsubsection*{Lax pair}

The next is to determine the associated Lax pair. 
By using the deformed current (\ref{dc-ga})\,, 
the Lax pair can be explicitly evaluated as 
\begin{eqnarray}
\label{lax-DHH}
\mL_\pm^{{\rm AT}}&=&-\partial_\pm\rho\frac{\lambda^{\pm1}}{2}\gamma^a_1
-\partial_\pm\theta\left[\frac{\la^{\pm1}}{2}\sinh\rho\,\gamma^a_3+\cosh\rho\,n^a_{13}\right] \nonumber\\
&& +i\partial_\pm\tau \left[\frac{\la^{\pm1}}{2}\cosh\rho\,\gamma^a_5
+\sinh\rho\,n^a_{15}\right]\nonumber\\
&& -\hat{G}\left(\partial_\pm\phi_1\mp\eta\sin^2\theta\sinh^2\rho\partial_\pm\,\phi_2\right) \nonumber \\
&& \qquad\qquad\qquad\times\left[\cos\theta\left(\frac{\la^{\pm1}}{2}\sinh\rho\,
\gamma^a_2+\cosh\rho\,n_{12}^a\right)+\sin\theta\,n_{23}^a\right] \nonumber\\
&& +i\hat{G}\left(\partial_\pm\phi_2\pm\eta\cos^2\theta\sinh^2\rho\partial_\pm\,\phi_1\right) \nonumber \\
&& \qquad\qquad\qquad\times\left[\sin\theta\left(\frac{\la^{\pm1}}{2}\sinh\rho\,\ga_0^a
-\cosh\rho\,n_{01}^a\right)-\cos\theta \,n^a_{03}\right]\,.
\end{eqnarray}
In the $\eta\to0$ limit, $\mL^{\rm{AT}}_\pm$ is reduced to the following form: 
\begin{eqnarray}
\label{lax-GAdS}
\mL^{\rm{GAdS}_5}_\pm&=&
-\partial_\pm\rho\frac{\lambda^{\pm1}}{2}\gamma^a_1
-\partial_\pm\theta\left[\frac{\la^{\pm1}}{2}\sinh\rho\,\gamma^a_3+\cosh\rho\,n^a_{13}\right]\nonumber\\
&& +i\partial_\pm\tau \left[\frac{\la^{\pm1}}{2}\cosh\rho\,\gamma^a_5
+\sinh\rho\,n^a_{15}\right] \nonumber \\
&& -\partial_\pm\phi_1
\left[\cos\theta\left(\frac{\la^{\pm1}}{2}\sinh\rho\,\gamma^a_2+\cosh\rho\,n_{12}^a\right)
+\sin\theta\,n_{23}^a\right] \nonumber \\
&& +i\partial_\pm\phi_2
\left[\sin\theta\left(\frac{\la^{\pm1}}{2}\sinh\rho\,\ga_0^a-\cosh\rho\,n_{01}^a\right)
-\cos\theta \,n^a_{03}\right]\,. 
\end{eqnarray}
This is nothing but a Lax pair for the global AdS$_5$\,.

\subsubsection*{Another derivation of Lax pair}

Even in this case, one can read off the replacement rules as well. 

\medskip 

The undeformed current is
\begin{eqnarray}
\label{udc-DHH}
A_{\pm}&=&-\partial_\pm\rho\frac{1}{2}\gamma^a_1
-\partial_\pm\theta\left[\frac{1}{2}\sinh\rho\,\gamma^a_3+\cosh\rho\,n^a_{13}\right] \nonumber \\
&& +i\partial_\pm\tau \left[\frac{1}{2}\cosh\rho\,\gamma^a_5
+\sinh\rho\,n^a_{15}\right] \nonumber \\
&& -\partial_\pm\phi_1
\left[\cos\theta\left(\frac{1}{2}\sinh\rho\,\gamma^a_2+\cosh\rho\,n_{12}^a\right)
+\sin\theta\,n_{23}^a\right] \nonumber \\
&& +i\partial_\pm\phi_2
\left[\sin\theta\left(\frac{1}{2}\sinh\rho\,\ga_0^a-\cosh\rho\,n_{01}^a\right)
-\cos\theta \,n^a_{03}\right]\,. \label{uc-ga} 
\end{eqnarray}
Then, by comparing the deformed current (\ref{dc-ga}) with the undeformed one (\ref{uc-ga})\,, 
the replacement rules are identified as 
\begin{eqnarray}
\partial_\pm \phi_1&&\longrightarrow~~\hat{G}
\left(\partial_\pm \phi_1\mp\eta\sin^2\theta\sinh^2\rho\,\partial_\pm\phi_2\right)\,,\nonumber\\
\partial_\pm \phi_2&&\longrightarrow~~\hat{G}
\left(\partial_\pm \phi_2\pm\eta\cos^2\theta\sinh^2\rho\,\partial_\pm\phi_1\right)\,.
\end{eqnarray}
By applying the replacement rules to the undeformed Lax pair (\ref{lax-GAdS})\,, 
one can reproduce the Lax pair (\ref{lax-DHH}) as well. 

\medskip 

Again, one can reinterpret the deformation as a twisted periodic boundary condition. 
After performing a similar analysis, the twisted boundary condition 
\begin{eqnarray}
\tilde{\phi}_1(\sigma=2\pi)&=&\tilde{\phi}_1(\sigma=0)+\frac{\eta}{\sqrt{\lambda_{\rm c}}}J_2+2\pi n_1\,, 
\nonumber\\
\tilde{\phi}_2(\sigma=2\pi)&=&\tilde{\phi}_2(\sigma=0)-\frac{\eta}{\sqrt{\lambda_{\rm c}}}J_1+2\pi n_2
\end{eqnarray}
with the undeformed AdS$_5\times$S$^5$ is equivalent to the deformed background 
with a periodic boundary condition. Here $J_i$ are Noether charges for rotation invariance 
in the $\phi_i$ directions. Integers $n_i$ are winding numbers along the $\phi_i$ directions.

\section{Conclusion and discussion}

We have explicitly derived Lax pairs for string theories on Yang-Baxter deformed backgrounds, 
1) gravity duals for NC gauge theories, 2) $\gamma$-deformations of S$^5$, 
3) Schr\"odinger spacetimes and 4) abelian twists of the global AdS$_5$\,. 
As another derivation, the Lax pair for gravity duals for NC gauge theories has been reproduced 
from the one for a $q$-deformed AdS$_5\times$S$^5$ by taking a scaling limit. 

\medskip 

As a byproduct, we have found a simple derivation of Lax pairs 
at least for all of the examples we have discussed here. 
After choosing a classical $r$-matrix and introducing a coordinate system, 
the replacement rules have been found out by comparing the deformed current $J$ with the undeformed current $A$\,. 
Then, by applying the rules to a Lax pair for the undeformed AdS$_5\times$S$^5$\,, 
one can construct the resulting Lax pair associated with the deformation. 
In addition, we have shown that each of the deformations considered here can be reinterpreted as 
a twisted periodic boundary condition with the undeformed AdS$_5\times$S$^5$\,, as in the work of \cite{Frolov}. 
It would be interesting to study the fermionic sector by following the work \cite{AAF}.  

\medskip 

This simple derivation really helps us to check the direct computation of Lax pairs based on Yang-Baxter deformations. 
In addition, it enables us to derive Lax pairs for Yang-Baxter deformations of Minkowski spacetime \cite{Minkowski}, 
for which the universal expression of Lax pair has not been obtained yet. Our procedure can play a significant role 
in studying along this direction. The result would be reported in another place \cite{future}. 

\medskip 

A more general question is what is the class of classical $r$-matrix for which one can deduce the replacement rule. 
Probably, it would be possible for some restricted $r$-matrices. This is also concerned with another question, 
what is the class of classical $r$-matrices for which the insertion of the operator can be eliminated 
by changing a boundary condition on the string world-sheet. 
It would be quite important to answer these questions. 

\medskip 

We believe that our concise prescription to construct Lax pairs would be helpful 
for further understanding of the gravity/CYBE correspondence.

\subsection*{Acknowledgments}

We are very grateful to Io Kawaguchi for collaboration at an early stage of this work. 
We also appreciate Heng-Yu Chen, Takuya Matsumoto and Stijn J.~van Tongeren for useful discussions. 
The work of T.K. is supported by the Japan Society for the Promotion of Science (JSPS).
The work of K.Y. is supported by Supporting Program for Interaction-based Initiative Team Studies 
(SPIRITS) from Kyoto University and by the JSPS Grant-in-Aid for Scientific Research (C) No.15K05051.
This work is also supported in part by the JSPS Japan-Russia Research Cooperative Program 
and the JSPS Japan-Hungary Research Cooperative Program.  

\appendix 

\section*{Appendix}

\section{Notation and convention}

We summarize here our notation and convention of the $\mathfrak{su}(2,2)$ and $\mathfrak{su}(4)$ generators.

\subsubsection*{The gamma matrices}

Let us first introduce the following gamma matrices: 
\begin{eqnarray}
\gamma_1=
\begin{pmatrix}
\;0~&~0~&~0~&-1\\
0&0&1&~0\\
0&1&0&~0\\
-1&0&0&~0\\
\end{pmatrix}\,, \quad 
\gamma_2=
\begin{pmatrix}
\;0~&~0~&~0~&~i\\
0&0&i&~0\\
0&-i&0&~0\\
-i&0&0&~0\\
\end{pmatrix}\,, \quad 
\gamma_3=
\begin{pmatrix}
\;0~&~0~&~1~&~0\\
0&0&0&~1\\
1&0&0&~0\\
0&1&0&~0\\
\end{pmatrix}\,,\nonumber
\end{eqnarray}
\begin{eqnarray}
\gamma_0=i\gamma_4=
\begin{pmatrix}
\;0~&~0~&1~&0\\
0&0&0&-1\\
-1&0&0&~0\\
0&1&0&~0\\
\end{pmatrix}\,, \quad 
\gamma_5=i\gamma_1\gamma_2\gamma_3\gamma_0=
\begin{pmatrix}
\;1~&~0~&~0~&0\\
0&1&0&~0\\
0&0&-1&~0\\
0&0&0&-1\\
\end{pmatrix}\,.
\end{eqnarray}
To embed $\mathfrak{su}(2,2)$ and $\mathfrak{su}(4)$ into $\mathfrak{su}(2,2|4)$\,,
we follow an $8\times8$ matrix representation as 
\begin{eqnarray}
&&\gamma_\mu^a=
\begin{pmatrix}
\gamma_\mu&\;0\\
0&\;0\\
\end{pmatrix}\,, \quad 
\gamma_5^a=
\begin{pmatrix}
\gamma_5&\;0\;\\
0&\;0\;\\
\end{pmatrix}\qquad\txt{with}\qquad\mu=0,1,2,3\,, 
\nonumber 
\\
&&\gamma_i^s=
\begin{pmatrix}
\;0&\;0\\
\;0&\;\gamma_i\\
\end{pmatrix}\,, \quad 
\gamma_5^s=
\begin{pmatrix}
\;0&\;0\\
\;0&\;\gamma_5\\
\end{pmatrix}\qquad\txt{with}\qquad\;  i=1,2,3,4\,.
\end{eqnarray}
Note that each block of the matrices is a $4\times4$ matrix.

\subsubsection*{The $\mathfrak{su}(2,2)$ and $\mathfrak{su}(4)$ generators}

The Lie algebras $\mathfrak{su}(2,2)\sim \mathfrak{so}(2,4)$ 
and $\mathfrak{su}(4) \sim \mathfrak{so}(6)$ are spanned as follows: 
\begin{eqnarray}
\alg{su}(2,2)&=&\text{span}_\mathbb{R}\{\,\ga_\mu^a\,,\,\ga_5^a\,,\,n^a_{\mu\nu}
=\frac{1}{4}[\ga_\mu^a\,,\ga_\nu^a]\,,\,n^a_{\mu 5}=\frac{1}{4}[\ga_\mu^a\,,\ga_5^a]~|~\mu\,,\nu=0,1,2,3\,\}\,, 
\label{so(2,4)} \nonumber \\
\alg{su}(4)&=&\text{span}_\mathbb{R}\{\,\ga_i^s\,,\,\ga_5^s\,,\,n^s_{ij}
=\frac{1}{4}[\ga_i^s\,,\ga_j^s]\,,\,n^s_{i5}=\frac{1}{4}[\ga_i^s\,,\ga_5^s] ~|~ i,j=1,2,3,4\,\}\,. 
\label{so(6)}
\end{eqnarray}
The subalgebras $\mathfrak{so}(1,4)$ and $\mathfrak{so}(5)$ in the spinor representation are formed as  
\begin{eqnarray}
\alg{so}(1,4)&=&\text{span}_\mathbb{R}\{\,n^a_{\mu\nu}\,,\,n^a_{\mu 5} ~|~ \mu\,,\nu=0,1,2,3\,\}\,,\nonumber\\
\alg{so}(5)&=&\text{span}_\mathbb{R}\{\, n^s_{ij}\,,\,n^s_{i5} ~|~i,j=1,2,3,4 \,\}\,.
\end{eqnarray}

\medskip

For a coset construction of Poincar$\acute{\txt{e}}$ AdS$_5$\,, it is useful to employ the following basis:  
\begin{eqnarray}
\alg{su}(2,2)&=&\text{span}_\mathbb{R}\{\,
p_\mu\,,k_\mu\,,h_1\,,h_2\,,h_3\,,n^a_{13}\,,n^a_{10}\,,n^a_{23}\,,n^a_{20}~|~\mu=0,1,2,3\,\}\,.
\end{eqnarray}
Here the generators $p_\mu$\,, $k_\mu$ and the Cartan generators $h_1\,,h_2\,,h_3$ are defined as
\begin{eqnarray}
p_\mu & \equiv&\frac{1}{2}\gamma_\mu^a-n_{\mu5}^a\,,\qquad 
k_\mu \equiv\frac{1}{2}\gamma_\mu^a+n_{\mu5}^a\,, \nonumber \\ 
h_1&\equiv& 2i\, n^a_{12}=\text{diag}(-1,1,-1,1,0,0,0,0)\,,\nonumber\\
h_2 &\equiv & 2n_{30}^a=\text{diag}(-1,1,1,-1,0,0,0,0)\,,\nonumber \\
h_3& \equiv &\ga^a_5=\text{diag}(1,1,-1,-1,0,0,0,0)\,. \nonumber
\label{ha}
\end{eqnarray}
Note that the generators $p_\mu$ and $k_\mu$ commute each other,
\begin{eqnarray}
[p_\mu\,,p_\nu]=[k_\mu\,,k_\nu]=[p_\mu\,,k_\nu]=0\qquad\txt{for}\qquad\mu\,,\nu=0,1,2,3\,.
\end{eqnarray} 

\medskip 

For the S$^5$ part, the Cartan generators $h_4\,,h_5\,,h_6$ of  $\mathfrak{su}(4)$ are given by
\begin{eqnarray}
h_4& \equiv &2i \,n^s_{12}=\text{diag}(0,0,0,0,-1,1,-1,1)
\,,\nonumber\\
h_5&\equiv &2i \,n^s_{34}=\text{diag}(0,0,0,0,-1,1,1,-1)
\,,\nonumber \\
h_6&\equiv &\ga^s_5=\text{diag}(0,0,0,0,1,1,-1,-1)
\,.\label{hs}
\end{eqnarray}
Since non-Cartan generators of $\mathfrak{su}(4)$ are not used in our analysis here, 
we will not write them down explicitly. 

\subsubsection*{The bosonic coset projectors}

In deriving the bosonic part of Lax pairs, it is necessary to employ the coset projectors 
$P_0$ and $P_2$ regarding the $\mathbb{Z}_2$-grading property. The projectors $P_0$ and $P_2$ 
are decomposed into the AdS$_5$ part and the S$^5$ part like 
\begin{eqnarray}
P_0(x) = P^a_0(x)+P^s_0(x)\,, \qquad P_2(x) = P^a_2(x)+P^s_2(x)\,, 
\end{eqnarray}
where $P^{a,s}_0$ and $P^{a,s}_2$ are the following coset projectors for $\mathfrak{so}(2,4)$ and $\mathfrak{su}(4)$\,,
\begin{eqnarray}
&&P_0^a\,:\quad \mathfrak{su}(2,2)~~ \longrightarrow~~ \mathfrak{so}(1,4)\,,\qquad P_2^a\,:
\quad \mathfrak{su}(2,2) ~~ \longrightarrow ~~ \frac{\mathfrak{su}(2,2)}{\mathfrak{so}(1,4)}\,, \nonumber \\
&&P_0^s\,:\quad \mathfrak{su}(4) ~~ \longrightarrow ~~ \mathfrak{so}(5)\,,\qquad\qquad P_2^s\,:
\quad \mathfrak{su}(4) ~~ \longrightarrow ~~ \frac{\mathfrak{su}(4)}{\mathfrak{so}(5)}\,. 
\end{eqnarray}
These coset projectors can be represented by the $\mathfrak{su}(2,2)$ and $\mathfrak{su}(4)$ generators 
as follows: 
\begin{eqnarray}
P^a_0(x)&=& \frac{1}{2}\sum_{\mu,\nu=0}^3
\frac{\Tr[n^a_{\mu\nu}x]}{\Tr[n^a_{\mu\nu}n^a_{\mu\nu}]}
n^a_{\mu\nu}+\sum_{\mu=0}^3\frac{\Tr[n^a_{\mu5}x]}{\Tr[n^a_{\mu5}n^a_{\mu5}]}n^a_{\mu5}\,,\nonumber \\
P^a_2(x)&=&\sum_{\mu=0}^3\frac{\Tr[\ga^a_\mu x]}{\Tr[\ga^a_\mu\ga^a_\mu]}\ga^a_\mu+
\frac{\Tr[\ga^a_5 x]}{\Tr[\ga^a_5\ga^a_5]}\ga^a_5\,,\nonumber \\
P_0^s(x)&=&\frac{1}{2}\sum_{\mu,\nu=1}^4\frac{\Tr[n^s_{\mu\nu}x]}{\Tr[n^s_{\mu\nu}n_{\mu\nu}]}
n^s_{\mu\nu}+\sum_{\mu=1}^4\frac{\Tr[n^s_{\mu5}x]}{\Tr[n^s_{\mu5}n^s_{\mu5}]}n^s_{\mu5}\,,\nonumber \\
P_2^s(x)&=&\sum_{\mu=1}^4\frac{\Tr[\ga^s_\mu x]}{\Tr[\ga^s_\mu\ga^s_\mu]}\ga^s_\mu+
\frac{\Tr[\ga^s_5 x]}{\Tr[\ga^s_5\ga^s_5]}\ga^s_5\,.
\end{eqnarray} 
The projectors are utilized in evaluating the deformed metric, NS-NS two-form and Lax pair.

\section{A Lax pair for a  $q$-deformed AdS$_5\times$S$^5$}

In this Appendix, let us consider a $q$-deformed AdS$_5\times$S$^5$ 
by employing the Yang-Baxter sigma model based on the mCYBE.
Then we explicitly present a Lax pair for a string theory on this background. 

\medskip

A typical skew-symmetric solution of the mCYBE is Drinfeld-Jimbo type \cite{DJ}. 
The classical action of the deformed AdS$_5\times$S$^5$ superstring associated with this $r$-matrix 
was constructed by Delduc-Magro-Vicedo \cite{DMV-string}.
The metric (in the string frame) and NS-NS two-form have been computed in \cite{ABF}. 
The deformed background is often called the $\eta$-deformed AdS$_5\times$S$^5$\,. 
Some specific limits\cite{HRT} and a mirror description\cite{mirror1,mirror2} have been studied. 
For various classical solutions, see\cite{KY-LL,magnon,AM-NR,Kame-coord,min-surf}.
Two-parameter generalizations have also been studied in\cite{HRT,bi-YB}.
For some arguments towards the complete supergravity solution, see\cite{LRT,ABF2,HT}.
More recently, another integrable deformation (called the $\lambda$-deformation) 
has been argued in \cite{Sfetsos,Hollowood}.
This deformation is closely related to the Yang-Baxter deformation 
by a Poisson-Lie duality \cite{Vicedo,Hoare-Tseytlin,Sfetsos,Klimcik-lambda}.

\subsection{Yang-Baxter deformations from the mCYBE}

Let us first give a short review on the  Yang-Baxter deformations of the AdS$_5\times$S$^5$ superstring
based on the mCYBE case \cite{DMV-string}.

\medskip

A $q$-deformed classical action of the AdS$_5\times$S$^5$ superstring \cite{DMV-string}
is given by 
\begin{eqnarray} 
S=-\frac{\sqrt{\lambda_{\rm c}}}{4}\,(1+\eta^2)\int_{-\infty}^\infty d\tau\int_0^{2\pi}d\sigma\,
(\ga^{\alpha\beta}-\epsilon^{\alpha\beta}){\rm STr}\Bigl[A_\alpha\, d\circ\frac{1}{1-\eta R_g\circ d}(A_\beta)\Bigr]\,,
\label{mYBsM}
\end{eqnarray}
The definition of $A_{\alpha}$ and $R_g$ is the same as in Section 2. 
A main difference is that the linear $R$-operator should satisfies the mCYBE 
\begin{eqnarray}
[R(X),R(Y)]-R([R(X),Y]+[X,R(Y)])=[X,Y]\,.
\label{mCYBE}
\end{eqnarray}
The projection operators $d$ is slightly different from the CYBE case like   
\begin{eqnarray}
d \equiv P_1+\frac{2}{1-\eta^2}P_2-P_3\,.
\label{d}
\end{eqnarray}
Namely, the coefficient in front of $P_2$ depends on $\eta$\,. This comes from the difference 
of the kappa transformation. 

\subsubsection*{The bosonic part of the Lagrangian}

We consider here the bosonic part of the deformed action (\ref{mYBsM})\,.  
The Lagrangian can be rewritten into a simple form, 
\begin{eqnarray}
L=\frac{\sqrt{\lambda_{\rm c}}}{2}\,\frac{1+\eta^2}{1-\eta^2}\,\text{STr}(A_-\,P_2(J_+))\,,
\label{Lq}
\end{eqnarray}
with the deformed current $J_\pm$ defined as 
\begin{eqnarray}
J_\pm \equiv \frac{1}{1\mp\varkappa R_g \circ P_2}\,A_{\pm}\,, \qquad \varkappa \equiv \frac{2\eta}{1-\eta^2}\,. 
\end{eqnarray}
The expression of $J_\pm$ is determined by solving the following equations: 
\begin{eqnarray}
\left(1\mp \varkappa  R_g  \circ  P_2\right)J_\pm&=&A_\pm\,. 
\label{J-mCYBE}
\end{eqnarray}
By taking a variation of (\ref{Lq})\,, the equation of motion is obtained as  
\begin{eqnarray}
\mathcal{E}\equiv\partial_+P_2(J_-)+\partial_-P_2(J_+)+[J_+,P_2(J_-)]+[J_-,P_2(J_+)]=0\,.
\label{eomq}
\end{eqnarray}
The undeformed current $A_\pm$ automatically satisfies the flatness condition 
\begin{eqnarray}
\mathcal{Z}\equiv\partial_+A_--\partial_-A_++[A_+,A_-]=0\,, 
\label{flatq}
\end{eqnarray}
which can be rewritten in terms of $J_\pm$ as
\begin{eqnarray}
\partial_+J_--\partial_-J_++[J_+,J_-]+\varkappa\, R_g(\E)
+\varkappa^2\,{\rm CYBE}_g(P_2(J_+),P_2(J_-))=0\,.
\label{flat2q}
\end{eqnarray}
Note that the quantity
\begin{eqnarray}
{\rm CYBE}_g(X,Y)\equiv[R_g(X),R_g(Y)]-R_g([R_g(X),Y]+[X,R_g(Y)]) 
\end{eqnarray}
results in $[X,Y]$\,, if the $R$-operator we are dealing with satisfies the mCYBE (\ref{mCYBE})\,. 
Thus, due to the mCYBE, the condition (\ref{flat2q}) is reduced to
\begin{eqnarray}
\mathcal{Z}=\partial_+J_--\partial_-J_++[J_+,J_-]+\varkappa\, R_g(\E)+\varkappa^2\,[P_2(J_+),P_2(J_-)]=0\,.
\label{flat3q}
\end{eqnarray}
In comparison to the CYBE case, the deformed  current $J_\pm$ no longer satisfies the flatness condition, 
even if the equation of motion (\ref{eomq}) is imposed.

\medskip

Finally, a Lax pair \cite{DMV-string} is given by 
\begin{eqnarray}
\mL_\pm=P_0(J_\pm)+\la^{\pm1}
\sqrt{1+\varkappa^2}\,P_2(J_\pm) 
\label{laxq}
\end{eqnarray}
with a spectral parameter $\lambda\in\mathbb{C}$\,. 
The flatness condition of $\mL_{\pm}$
\begin{eqnarray}
\partial_+\mL_--\partial_-\mL_-+[\mL_+,\mL_-]=0 
\end{eqnarray}
leads to the equation of motion (\ref{eomq}) and the flatness condition (\ref{flat2q})\,.

\subsection{A Lax pair for a $q$-deformed AdS$_5\times$S$^5$}

We shall study the bosonic part with a classical $r$-matrix of Drinfeld-Jimbo type \cite{DJ},  
\begin{eqnarray}
r_{\txt{DJ}}=-i\sum_{a<b}E_{ab}\wedge E_{ba}-i\sum_{c<d}E_{cd}\wedge E_{dc}\,, 
\end{eqnarray}
where $E_{ab}~(a,b=1,\ldots,4)$ and $E_{cd}~(c,d=5,\ldots,8)$ are 
the fundamental representations of $\mathfrak{su}(2,2)$ and $\mathfrak{su}(4)$\,, respectively. 
This is a solution of the mCYBE (\ref{mCYBE})\,.

\medskip

To construct the bosonic part of the deformed AdS$_5\times$S$^5$ with the global coordinates, 
we will work with  a bosonic  element represented by 
\begin{eqnarray}
g=g_a\cdot g_s ~~\in ~SU(2,2)\times SU(4)\,,
\end{eqnarray} 
where the group elements of $SU(2,2)$ and $SU(4)$ are parameterized as, respectively, 
\begin{eqnarray}
g_a(\tau,\sigma)&=&\exp\left[\frac{i}{2}\sum_{i=1}^3\psi_i h_i\right]
\exp[-\zeta\, n^a_{13}]\exp\Bigl[-\frac{1}{2}\,\rho\, \ga^a_1\Bigr]\,,\nonumber\\
g_s(\tau,\sigma)&=&\exp\left[\frac{i}{2}\sum_{i=1}^3\phi_i h_{i+3}\right]
\exp[-\xi \,n^s_{13}]\exp\Bigl[-\frac{i}{2}\,r\, \ga^s_1\Bigr]\,.
\end{eqnarray}
The AdS$_5$ part is described by the coordinates $\psi_3\,(\equiv t)\,,\psi_1\,,\psi_2\,, \zeta\,, \rho$\,. 
The S$^5$ part is parameterized by the angle variables $\phi_1\,,\phi_2\,,\phi_3\,, \xi\,,r$\,.

\medskip 

In the present case, the deformed current $J_{\pm}$ is decomposed into two pieces: $J_\pm=J_\pm^a+J_\pm^s$\,. 
Then, by solving the equations in (\ref{J-mCYBE})\,, $J_{\pm}^a$ and $J_{\pm}^s$ are determined as  
\begin{eqnarray}
J_\pm^a&=&-f_a(\rho)\partial_\pm\rho\left[
\frac{1}{2}(\ga_1^a\pm i\varkappa\sinh\rho\,\ga_5^a)
\pm i\varkappa\cosh\rho\,n_{15}^a\right]\nonumber \\
&&+f_a(\rho)\partial_\pm t\left[
\frac{1}{2}\cosh\rho\,(i\ga_5^a\pm\varkappa\sinh\rho\,\ga_1^a)
+i(1+\varkappa^2)\sinh\rho\,n_{15}^a\right]\nonumber \\
&&-g_a(\rho\,,\zeta)\partial_\pm\zeta\biggl[\frac{1}{2}\sinh\rho
\left(\ga_3^a\pm\varkappa\sin\zeta \sinh^2\rho\ga_2^a\right)\nonumber \\
&&\qquad\qquad\qquad\qquad\pm i\varkappa\sinh\rho\cosh\rho
\left(n_{35}^a\pm\varkappa\sin\zeta\sinh^2\rho\,n_{25}^a\right)
\mp\varkappa \cos\zeta\sinh^2\rho\,n_{23}^a\nonumber \\
&&\qquad\qquad\qquad\qquad+\cosh\rho
\left(n_{13}^a\pm\varkappa\sin\zeta\sinh ^2\rho\,n_{12}^a\right)
\biggr]\nonumber \\
&&-g_a(\rho\,,\zeta)\partial_\pm\psi_1
\biggl[\frac{1}{2}
\cos\zeta\sinh\rho(\ga_2^a\mp\varkappa\sin\zeta\sinh^2\rho\,\ga_3^a) \nonumber \\
&&\qquad\qquad\qquad\qquad+\sin\zeta(1+\varkappa^2\sinh^4\rho)\,n_{23}^a 
+\cos\zeta\cosh\rho\,(n_{12}^a\mp\varkappa\sin\zeta\sinh^2\rho\,n_{13}^a) \nonumber \\
&&\qquad\qquad\qquad\qquad\pm i\varkappa\cos\zeta\sinh\rho
\cosh\rho(n_{25}^a\mp\varkappa\sin\zeta\sinh^2\rho\,n_{35}^a)
\biggr]\nonumber \\
&&+\partial_\pm\psi_2
\biggl[i\sin\zeta\sinh\rho\left(\frac{1}{2}
\ga_0^a
\pm i\varkappa\cosh\rho\,n_{05}^a\right)\nonumber \\
&&\qquad\qquad\qquad-i\sin\zeta\cosh\rho\,n_{01}^a
- i\cos\zeta(n_{03}^a\mp\varkappa\sin\zeta\sinh^2\rho\,n_{02}^a) \biggr]
\,, \label{def-mCYBE1} \\
J_\pm^s &=& -f_s(r)\partial_\pm r\left[i\frac{1}{2}(\ga_1^s\mp\varkappa\sin r\,\ga_5^s) 
\mp\varkappa\cos r\, n_{15}^s\right]\nonumber \\
&& -g_s(r,\xi)\partial_\pm \xi\Bigl[i\frac{1}{2}\sin r
\left(\ga_3^s\mp\varkappa\sin\xi\sin^2r\,\ga_2^s\right)\nonumber \\
&& \qquad\qquad\qquad\pm\varkappa\cos\xi\sin^2r\,n_{23}^s\mp\varkappa\sin r\cos r\left(n_{35}^s
\mp\varkappa\sin\xi\sin^2r\,n_{25}^s \right)\nonumber \\
&&\qquad\qquad\qquad+\cos r\left(n_{13}^s\mp\varkappa\sin\xi\sin^2r\, n_{12}^s\right)\Bigr]\nonumber \\
&& -g_s(r,\xi)\partial_\pm\phi_1\biggl[i\frac{1}{2}\sin r\cos\xi
\left(\ga_2^s\pm\varkappa\sin\xi\sin^2r\,\ga_3 ^s\right)\nonumber \\
&& \qquad\qquad\qquad+\sin\xi\left(1+\varkappa^2\sin^4r\right)n_{23}^s
\mp\varkappa\sin r\cos r\cos\xi\left(n_{25}^s\pm\varkappa\sin\xi\sin^2r\,n_{35}^s\right)\nonumber \\
&& \qquad\qquad\qquad+\cos r\cos\xi\left(n^s_{12}\pm\varkappa \sin^2r\sin\xi\,n_{13}^s\right)
\biggr]\nonumber \\
&& -\partial_\pm \phi_2\biggl[\sin r\sin\xi\left(i
\frac{1}{2}\,\ga_4^s\mp\varkappa\cos r\,n_{45}^s\right)\nonumber \\
&& \qquad\qquad\qquad+\sin\xi\sin r\cos r\,n_{14}^s 
+\cos\xi(n_{34}^s\pm\varkappa \sin\xi \sin^2r\,n_{24}^s)\biggr]\nonumber \\
&& +f_s(r)\partial_\pm \phi_3 \Bigl[i\frac{1}{2}\cos r\,(\ga_5^s\pm\varkappa\sin r\,\ga_1^s)
-(1+\varkappa^2)\sin r\,n_{15}^s\Bigr]\,, \label{def-mCYBE2}
\end{eqnarray}
where we have introduced new functions defined as 
\begin{eqnarray}
f_a(\rho) &\equiv& \frac{1}{1-\varkappa^2\sinh^2\rho}\,,\quad\quad 
g_a(\rho,\zeta) \equiv \frac{1}{1+\varkappa^2\sin^2\zeta\sinh^4\rho}\,,\nonumber \\
f_s(r) &\equiv & \frac{1}{1+\varkappa^2\sin^2r}\,,\quad\quad 
g_s(r,\xi) \equiv \frac{1}{1+\varkappa^2 \sin^2\xi\sin^4r}\,.
\end{eqnarray} 
The deformed currents in (\ref{def-mCYBE1}) and  (\ref{def-mCYBE2}) enable us 
to compute (i) the metric and NS-NS two-form and (ii) the explicit form of the Lax pair. 

\medskip 
 
Firstly, the resulting metric and NS-NS two-form are given by \cite{ABF} 
\begin{eqnarray}
ds_{\txt{AdS}_5}^2&=&
\sqrt{1+\varkappa^2}\,\biggl[\frac{1}{1-\varkappa^2\sinh^2\rho}\left(-\cosh ^2\rho\, dt^2+d\rho^2\right)\nonumber \\
\qquad\qquad&&+\frac{1}{1+\varkappa^2\sin^2\zeta\sinh^4\rho}\sinh^2\rho\left(d\zeta^2+\cos ^2\zeta\,  
(d\psi_1)^2\right)+ \sinh ^2\rho \sin ^2\zeta\,( d\psi_2)^2\biggr]\,,\nonumber \\
B_{\txt{AdS}_5}&=&\varkappa\sqrt{1+\varkappa^2}\, 
\frac{\sinh ^4\rho\sin 2 \zeta}{1+\varkappa^2\sin^2\zeta\sinh^4\rho} \,d\psi_1\wedge d\zeta\,,
\label{q-Ads5}\\\nonumber\\
ds^2_{\text{S}^5}&=&
\sqrt{1+\varkappa^2}\,\biggl[\frac{1}{1+\varkappa^2\sin^2r}
\left(\cos ^2r(d\phi_3)^2+dr^2\right)\nonumber \\&&\hspace{1.5cm}+
\frac{\sin^2r}{1+\varkappa^2\sin^2\xi\sin^4r}
\left(d\xi ^2 +\cos^2\xi\, (d\phi_1)^2\right)+\sin^2r\sin^2\xi \,(d\phi_2)^2\biggr]\,,\nonumber \\
B_{\text{S}^5}&=&\varkappa\sqrt{1+\varkappa^2}\, 
\frac{\sin^4r\sin2\xi}{1+\varkappa^2\sin^2\xi\sin^4r} \,d\phi_1\wedge d\xi\,.
\label{q-s5}
\end{eqnarray}
Here total derivative terms in the NS-NS two-form have been ignored.

\medskip

Secondly, the Lax pair (\ref{laxq}) is also decomposed into two parts: 
$\mathcal{L}_{\pm} = \mathcal{L}_{\pm}^a + \mathcal{L}_{\pm}^s$\,.  
Then the explicit forms of $\mathcal{L}_{\pm}^a$ and $\mathcal{L}_{\pm}^s$ turn out to be
\begin{eqnarray}
\mL_\pm^a&=&-f_a(\rho)\partial_\pm\rho\left[
\frac{\la^{\pm1}}{2}\sqrt{1+\varkappa^2}(\ga_1^a\pm i\varkappa\sinh\rho\,\ga_5^a)
\pm i\varkappa\cosh\rho\,n_{15}^a\right]\nonumber \\
&&+f_a(\rho)\partial_\pm t\left[
\frac{\la^{\pm1}}{2}\sqrt{1+\varkappa^2}\cosh\rho\,
(i\ga_5^a\pm\varkappa\sinh\rho\,\ga_1^a)
+i(1+\varkappa^2)\sinh\rho\,n_{15}^a\right]\nonumber \\
&&-g_a(\rho\,,\zeta)\partial_\pm\zeta\biggl[\frac{\la^{\pm1}}{2}\sqrt{1+\varkappa^2}
\sinh\rho\left(\ga_3^a\pm\varkappa\sin\zeta \sinh^2\rho\ga_2^a\right)\nonumber \\
&&\qquad\qquad\qquad\qquad\pm i\varkappa\sinh\rho\cosh\rho\left(n_{35}^a
\pm\varkappa\sin\zeta\sinh^2\rho\,n_{25}^a\right)
\mp\varkappa \cos\zeta\sinh^2\rho\,n_{23}^a\nonumber \\
&&\qquad\qquad\qquad\qquad+\cosh\rho
\left(n_{13}^a\pm\varkappa\sin\zeta\sinh ^2\rho\,n_{12}^a\right)
\biggr]\nonumber \\
&&-g_a(\rho\,,\zeta)\partial_\pm\psi_1
\biggl[\frac{\la^{\pm1}}{2}
\sqrt{1+\varkappa^2}\cos\zeta\sinh\rho 
(\ga_2^a\mp\varkappa\sin\zeta\sinh^2\rho\,\ga_3^a) \nonumber \\
&&\qquad\qquad\qquad\qquad+\sin\zeta(1+\varkappa^2\sinh^4\rho)\,n_{23}^a 
+\cos\zeta\cosh\rho\,(n_{12}^a\mp\varkappa\sin\zeta\sinh^2\rho\,n_{13}^a) \nonumber \\
&&\qquad\qquad\qquad\qquad\pm i\varkappa\cos\zeta\sinh\rho
\cosh\rho(n_{25}^a\mp\varkappa\sin\zeta\sinh^2\rho n_{35}^a)
\biggr]\nonumber \\
&&+\partial_\pm\psi_2
\biggl[i\sin\zeta\sinh\rho\left(\frac{\la^{\pm1}}{2}
\sqrt{1+\varkappa^2}\ga_0^a
\pm i\varkappa\cosh\rho\,n_{05}^a\right)\nonumber \\
&&\qquad\qquad\qquad-i\sin\zeta\cosh\rho\,n_{01}^a- i\cos\zeta(n_{03}^a
\mp\varkappa\sin\zeta\sinh^2\rho\,n_{02}^a) \biggr]
\,, \label{q-lax} \\
\mL_\pm^s &=&
-f_s(r)\partial_\pm r\left[i\frac{\la^{\pm1}}{2}
\sqrt{1+\varkappa^2}(\ga_1^s\mp\varkappa\sin r\,\ga_5^s) 
\mp\varkappa\cos r\, n_{15}^s\right]\nonumber \\
&&-g_s(r,\xi)\partial_\pm \xi\Bigl[i\frac{\la^{\pm1}}{2}\sqrt{1+\varkappa^2}
\sin r\left(\ga_3^s\mp\varkappa\sin\xi\sin^2r\,\ga_2^s\right)\nonumber \\
&&\qquad\qquad\qquad\pm\varkappa\cos\xi\sin^2r\,n_{23}^s\mp\varkappa\sin r\cos r
\left(n_{35}^s\mp\varkappa\sin\xi\sin^2r\,n_{25}^s \right)\nonumber \\
&&\qquad\qquad\qquad+\cos r\left(n_{13}^s\mp\varkappa\sin\xi\sin^2r\, n_{12}^s\right)
\Bigr]\nonumber \\
&&-g_s(r,\xi)\partial_\pm \phi_1\biggl[i\frac{\la^{\pm1}}{2}\sqrt{1+\varkappa^2}
\sin r\cos\xi\left(\ga_2^s\pm\varkappa\sin\xi\sin^2r\,\ga_3 ^s\right)\nonumber \\
&&\qquad\qquad\qquad+\sin\xi\left(1+\varkappa^2\sin^4r\right)n_{23}^s
\mp\varkappa\sin r\cos r\cos\xi\left(n_{25}^s\pm\varkappa\sin\xi\sin^2r\,n_{35}^s\right)\nonumber \\
&&\qquad\qquad\qquad+\cos r\cos\xi\left(n^s_{12}\pm\varkappa \sin^2r\sin\xi\,n_{13}^s\right)
\biggr]\nonumber \\
&& -\partial_\pm \phi_2\biggl[\sin r\sin\xi\left(i\frac{\la^{\pm1}}{2}\sqrt{1+\varkappa^2}\,
\ga_4^s\mp\varkappa\cos r\,n_{45}^s\right)\nonumber \\
&& \qquad\qquad\qquad +\sin\xi\sin r\cos r\,n_{14}^s 
+\cos\xi(n_{34}^s\pm\varkappa \sin\xi \sin^2r\,n_{24}^s)\biggr]\nonumber \\
&& +f_s(r)\partial_\pm \phi_3
\Bigl[i\frac{\la^{\pm1}}{2}\sqrt{1+\varkappa^2}\cos r\,(\ga_5^s\pm\varkappa\sin r\,\ga_1^s)
-(1+\varkappa^2)\sin r\,n_{15}^s \Bigr]\,. 
\label{q-lax2}
\end{eqnarray}
The expression (\ref{q-lax}) is utilized in Section 3 in order to reproduce the desired Lax pair 
as a scaling limit.


\begin{thebibliography}{99}

\bibitem{M}  
  J.~M.~Maldacena,
  ``The large N limit of superconformal field theories and supergravity,''
  Int.\ J.\ Theor.\ Phys.\  {\bf 38} (1999) 1113
   [Adv.\ Theor.\ Math.\ Phys.\  {\bf 2} (1998) 231]
  [hep-th/9711200].

\bibitem{review}
  N.~Beisert {\it et al.},
  ``Review of AdS/CFT Integrability: An Overview,''
  Lett.\ Math.\ Phys.\  {\bf 99} (2012) 3
  [arXiv:1012.3982 [hep-th]].

\bibitem{MT}
  R.~R.~Metsaev and A.~A.~Tseytlin,
  ``Type IIB superstring action in AdS$_5\times$S$^5$ background,''  
  Nucl.\ Phys.\ B {\bf 533} (1998) 109
  [hep-th/9805028].

\bibitem{BPR}
  I.~Bena, J.~Polchinski and R.~Roiban,
  ``Hidden symmetries of the AdS$_5\times$S$^5$ superstring,''
  Phys.\ Rev.\ D {\bf 69} (2004) 046002
  [hep-th/0305116].
  

\bibitem{AF-review}
  G.~Arutyunov and S.~Frolov,
  ``Foundations of the AdS$_5\times$S$^5$ Superstring. Part I,''
  J.\ Phys.\ A {\bf 42} (2009) 254003
  [arXiv:0901.4937 [hep-th]].        

\bibitem{Klimcik}
 C.~Klimcik,
  ``Yang-Baxter sigma models and dS/AdS T duality,'' JHEP {\bf 0212} (2002)
051 [hep-th/0210095];   ``On integrability of the Yang-Baxter sigma-model,'' J.\ Math.\
Phys.\ {\bf 50} (2009) 043508 [arXiv:0802.3518 [hep-th]]. 

\bibitem{DMV}
  F.~Delduc, M.~Magro and B.~Vicedo,
  ``On classical $q$-deformations of integrable $\sigma$-models,'' 
JHEP {\bf 1311} (2013) 192 [arXiv:1308.3581 [hep-th]].     
  
\bibitem{KYhybrid}
  I.~Kawaguchi and K.~Yoshida,
  ``Hybrid classical integrability in squashed sigma models,''
  Phys.\ Lett.\ B\ {\bf 705} (2011) 251
  [arXiv:1107.3662 [hep-th]]; 
   ``Hybrid classical integrable structure of squashed sigma models: A short summary,''  
  J.\ Phys.\ Conf.\ Ser.\  {\bf 343} (2012) 012055 
  [arXiv:1110.6748 [hep-th]]; 
  ``Hidden Yangian symmetry in sigma model on squashed sphere,''
  JHEP {\bf 1011} (2010) 032. 
  [arXiv:1008.0776 [hep-th]].     

\bibitem{KMY-QAA}
  I.~Kawaguchi, T.~Matsumoto and K.~Yoshida,
  ``The classical origin of quantum affine algebra in squashed sigma models,''  
  JHEP {\bf 1204} (2012) 115  [arXiv:1201.3058 [hep-th]];  
  ``On the classical equivalence of monodromy matrices in squashed sigma model,''  
  JHEP {\bf 1206} (2012) 082  [arXiv:1203.3400 [hep-th]].

\bibitem{KOY}
  I.~Kawaguchi, D.~Orlando and K.~Yoshida,
  ``Yangian symmetry in deformed WZNW models on squashed spheres,''
  Phys.\ Lett.\  B {\bf 701} (2011) 475. 
  [arXiv:1104.0738 [hep-th]]; I.~Kawaguchi and K.~Yoshida,
   ``A deformation of quantum affine algebra in squashed
 WZNW models,'' J.\ Math.\ Phys.\ {\bf 55} (2014) 062302 [arXiv:1311.4696 [hep-th]].

\bibitem{Sch}
  I.~Kawaguchi and K.~Yoshida,
  ``Classical integrability of Schrodinger sigma models and $q$-deformed Poincare symmetry,''  
JHEP {\bf 1111} (2011) 094  [arXiv:1109.0872 [hep-th]]; 
  ``Exotic symmetry and monodromy equivalence in Schrodinger sigma models,''  
JHEP {\bf 1302} (2013) 024  [arXiv:1209.4147 [hep-th]]; 
  I.~Kawaguchi, T.~Matsumoto and K.~Yoshida,
  ``Schroedinger sigma models and Jordanian twists,''  
JHEP {\bf 1308} (2013) 013  [arXiv:1305.6556 [hep-th]].  

\bibitem{DMV-string} 
 F.~Delduc, M.~Magro and B.~Vicedo,
  ``An integrable deformation of the AdS$_5\times$S$^5$ superstring
action,'' Phys.\ Rev.\ Lett.\  {\bf 112} (2014) 051601
  [arXiv:1309.5850 [hep-th]];
 F.~Delduc, M.~Magro and B.~Vicedo,
  ``Derivation of the action and symmetries of the $q$-deformed AdS$_5 \times $S$^5$ superstring,''
  JHEP {\bf 1410}, 132 (2014)
  [arXiv:1406.6286 [hep-th]].

\bibitem{KMY-JordanianAdSxS}
 I.~Kawaguchi, T.~Matsumoto and K.~Yoshida,
  ``Jordanian deformations of the AdS$_5\times$S$^5$ superstring,'' 
  JHEP {\bf 1404} (2014) 153
  [arXiv:1401.4855 [hep-th]]; 

\bibitem{KMY-SUGRA} 
 I.~Kawaguchi, T.~Matsumoto and K.~Yoshida,
  ``A Jordanian deformation of AdS space in type IIB supergravity,''
  JHEP {\bf 1406} (2014) 146
  [arXiv:1402.6147 [hep-th]]. 
  
\bibitem{MY1}
T.~Matsumoto and K.~Yoshida,
  ``Lunin-Maldacena backgrounds from the classical Yang-Baxter equation 
-- Towards the gravity/CYBE correspondence,''
  JHEP {\bf 1406} (2014) 135
  [arXiv:1404.1838 [hep-th]]. 

\bibitem{MY2}
 T.~Matsumoto and K.~Yoshida,
  ``Integrability of classical strings dual for noncommutative gauge theories,''
  JHEP {\bf 1406} (2014) 163
  [arXiv:1404.3657 [hep-th]].

\bibitem{YB1}
  T.~Matsumoto and K.~Yoshida,
  ``Yang-Baxter deformations and string dualities,''
  JHEP {\bf 1503} (2015) 137
  [arXiv:1412.3658 [hep-th]].

\bibitem{YB2}
  T.~Matsumoto and K.~Yoshida,
  ``Yang-Baxter sigma models based on the CYBE,''
  Nucl.\ Phys.\ B {\bf 893} (2015) 287
  [arXiv:1501.03665 [hep-th]].

\bibitem{Sch-YB}
  T.~Matsumoto and K.~Yoshida,
  ``Schr\"odinger geometries arising from Yang-Baxter deformations,''
  JHEP {\bf 1504} (2015) 180
  [arXiv:1502.00740 [hep-th]].

\bibitem{Stijn}
  S.~J.~van Tongeren,
  ``On classical Yang-Baxter based deformations of the AdS$_{5}\times$S$^{5}$ superstring,''
  JHEP {\bf 1506} (2015) 048
  [arXiv:1504.05516 [hep-th]];
 S.~J.~van Tongeren,
  ``Yang-Baxter deformations, AdS/CFT, and twist-noncommutative gauge theory,''
  arXiv:1506.01023 [hep-th].

\bibitem{LM}
 O.~Lunin and J.~M.~Maldacena,
  ``Deforming field theories with $U(1) \times U(1)$ global symmetry and their gravity duals,''  
JHEP {\bf 0505} (2005) 033  [hep-th/0502086].

\bibitem{Frolov}
  S.~Frolov,
  ``Lax pair for strings in Lunin-Maldacena background,''
  JHEP {\bf 0505} (2005) 069
  [hep-th/0503201].

\bibitem{HI}
  A.~Hashimoto and N.~Itzhaki,
  ``Noncommutative Yang-Mills and the AdS / CFT correspondence,''
  Phys.\ Lett.\ B {\bf 465} (1999) 142
  [hep-th/9907166]. \\
  J.~M.~Maldacena and J.~G.~Russo,
  ``Large N limit of noncommutative gauge theories,''
  JHEP {\bf 9909} (1999) 025
  [hep-th/9908134].

\bibitem{10DSch}
C.~P.~Herzog, M.~Rangamani and S.~F.~Ross,
  ``Heating up Galilean holography,''
  JHEP {\bf 0811} (2008) 080
  [arXiv:0807.1099 [hep-th]]; 
 J.~Maldacena, D.~Martelli and Y.~Tachikawa,
  ``Comments on string theory backgrounds with non-relativistic conformal symmetry,''
  JHEP {\bf 0810} (2008) 072
  [arXiv:0807.1100 [hep-th]]; 
A.~Adams, K.~Balasubramanian and J.~McGreevy,
  ``Hot Spacetimes for Cold Atoms,''
  JHEP {\bf 0811} (2008) 059
  [arXiv:0807.1111 [hep-th]]. 

\bibitem{CYBE}
  T.~Matsumoto and K.~Yoshida,
  ``Integrable deformations of the AdS$_5\times$S$^5$ superstring 
and the classical Yang-Baxter equation 
-- Towards the gravity/CYBE correspondence --,''
  J.\ Phys.\ Conf.\ Ser.\  {\bf 563} (2014) 1,  012020
  [arXiv:1410.0575 [hep-th]].
  
\bibitem{LLM}
 H.~Lin, O.~Lunin and J.~M.~Maldacena,
  ``Bubbling AdS space and 1/2 BPS geometries,''
  JHEP {\bf 0410} (2004) 025
  [hep-th/0409174].  
  
\bibitem{Minkowski}
  T.~Matsumoto, D.~Orlando, S.~Reffert, J.~Sakamoto and K.~Yoshida,
  ``Yang-Baxter deformations of Minkowski spacetime,''
  arXiv:1505.04553 [hep-th].  
  
\bibitem{kappa}
S.~Zakrzewski, ``Poisson structures on the Poincar\'e group,'' Comm.\ Math.\ Phys. {\bf 185} (1997) 285 
[q-alg/9602001]; 
J.~Lukierski and M.~Woronowicz,
  ``New Lie-algebraic and quadratic deformations of Minkowski space from twisted Poincare symmetries,''
  Phys.\ Lett.\ B {\bf 633} (2006) 116
  [hep-th/0508083]; 
 A.~Borowiec, J.~Lukierski, M.~n.~Mozrzymas and V.~N.~Tolstoy,
  ``N=1/2 Deformations of Chiral Superspaces from New Twisted Poincare and Euclidean Superalgebras,''
  JHEP {\bf 1206} (2012) 154
  [arXiv:1112.1936 [hep-th]].  

\bibitem{r-dS}
  A.~Ballesteros, F.~J.~Herranz and N.~R.~Bruno,
  ``Quantum (anti)de Sitter algebras and generalizations of the kappa-Minkowski space,''
  hep-th/0409295.

\bibitem{KW}
  I.~R.~Klebanov and E.~Witten,
  ``Superconformal field theory on three-branes at a Calabi-Yau singularity,''
  Nucl.\ Phys.\ B {\bf 536} (1998) 199
  [hep-th/9807080].

\bibitem{BZ}
P.~Basu and L.~A.~Pando Zayas,
  ``Chaos Rules out Integrability of Strings in AdS$_5 \times T^{1,1}$,''  
Phys.\ Lett.\ B {\bf 700} (2011) 243 [arXiv:1103.4107 [hep-th]]; 
  ``Analytic Non-integrability in String Theory,''
  Phys.\ Rev.\ D {\bf 84} (2011) 046006
  [arXiv:1105.2540 [hep-th]].

\bibitem{Penrose-T11}
  Y.~Asano, D.~Kawai, H.~Kyono and K.~Yoshida,
  ``Chaotic strings in a near Penrose limit of AdS$_5\times T^{1,1}$,''
 JHEP {\bf 1508} (2015) 060 [arXiv:1505.07583 [hep-th]].
 
 \bibitem{CO}
  A.~Catal-Ozer,
  ``Lunin-Maldacena deformations with three parameters,''
  JHEP {\bf 0602} (2006) 026
  [hep-th/0512290].
 
\bibitem{CMY-T11}
  P.~M.~Crichigno, T.~Matsumoto and K.~Yoshida,
  ``Deformations of $T^{1,1}$ as Yang-Baxter sigma models,''
  JHEP {\bf 1412} (2014) 085
  [arXiv:1406.2249 [hep-th]]. 
 
\bibitem{MS}  
D.~Dhokarh, S.~S.~Haque and A.~Hashimoto,
  ``Melvin Twists of global AdS$_5\times$S$^5$ and their Non-Commutative Field Theory Dual,''
  JHEP {\bf 0808} (2008) 084
  [arXiv:0801.3812 [hep-th]]; T.~McLoughlin and I.~Swanson,
  ``Integrable twists in AdS/CFT,''
  JHEP {\bf 0608} (2006) 084
  [hep-th/0605018]. 
 
\bibitem{ABF} 
 G.~Arutyunov, R.~Borsato and S.~Frolov,
  ``S-matrix for strings on $\eta$-deformed AdS$_5\times$S$^5$,'' 
  JHEP {\bf 1404} (2014) 002
  [arXiv:1312.3542 [hep-th]].
 
\bibitem{ABF2}
  G.~Arutyunov, R.~Borsato and S.~Frolov,
  ``Puzzles of eta-deformed  $AdS_5\times S^5$,''
  arXiv:1507.04239 [hep-th].
  
\bibitem{DJ}
  V.~G.~Drinfel'd,
  ``Hopf algebras and the quantum Yang-Baxter equation,'' 
  Sov.\ Math.\ Dokl.\ {\bf 32} (1985) 254; 
  ``Quantum groups,''
  J.\ Sov.\ Math.\  {\bf 41} (1988) 898 
  [Zap.\ Nauchn.\ Semin.\  {\bf 155}, 18 (1986)]; 
  M.~Jimbo,
  ``A $q$ difference analog of $U(g)$ and the Yang-Baxter equation,''
  Lett.\ Math.\ Phys.\  {\bf 10} (1985) 63.        
  
\bibitem{LS}  
  R.~G.~Leigh and M.~J.~Strassler,
  ``Exactly marginal operators and duality in four-dimensional N=1 supersymmetric gauge theory,''
  Nucl.\ Phys.\  B {\bf 447} (1995) 95
  [arXiv:hep-th/9503121]. 
    
\bibitem{AAF}
  L.~F.~Alday, G.~Arutyunov and S.~Frolov,
  ``Green-Schwarz strings in TsT-transformed backgrounds,''
  JHEP {\bf 0606} (2006) 018
  [hep-th/0512253].


\bibitem{SYY}
 S.~Schafer-Nameki, M.~Yamazaki and K.~Yoshida,
  ``Coset Construction for Duals of Non-relativistic CFTs,''
  JHEP {\bf 0905} (2009) 038
  [arXiv:0903.4245 [hep-th]].

\bibitem{future}
  A.~Borowiec, H.~Kyono, J.~Lukierski, J.~Sakamoto and K.~Yoshida,
  ``Yang-Baxter sigma models and Lax pairs arising from $\kappa$-Poincar\'e $r$-matrices,''
  arXiv:1510.03083 [hep-th].

\bibitem{HRT}
  B.~Hoare, R.~Roiban and A.~A.~Tseytlin,
  ``On deformations of AdS$_n\times$S$^n$ supercosets,''
  JHEP {\bf 1406} (2014) 002
  [arXiv:1403.5517 [hep-th]].

\bibitem{mirror1}  
 G.~Arutynov, M.~de Leeuw and S.~J.~van Tongeren,
  ``The exact spectrum and mirror duality of the (AdS$_{5} \times$ S$^5$)$_{\eta}$ superstring,''
  Theor.\ Math.\ Phys.\  {\bf 182} (2015) 1,  23
   [Teor.\ Mat.\ Fiz.\  {\bf 182} (2014) 1,  28]
  [arXiv:1403.6104 [hep-th]];

\bibitem{mirror2}  
  G.~Arutyunov and S.~J.~van Tongeren,
  ``$\mathrm{AdS}_5 \times \mathrm{S}^5$ mirror model as a string,''
  Phys.\ Rev.\ Lett.\  {\bf 113} (2014) 261605
  [arXiv:1406.2304 [hep-th]]; 
  ``Double Wick rotating Green-Schwarz strings,''
  JHEP {\bf 1505} (2015) 027
  [arXiv:1412.5137 [hep-th]].

\bibitem{KY-LL} 
  T.~Kameyama and K.~Yoshida,
  ``Anisotropic Landau-Lifshitz sigma models from $q$-deformed AdS$_5 \times $S$^5$ superstrings,''
  JHEP {\bf 1408} (2014) 110
  [arXiv:1405.4467 [hep-th]];
  ``String theories on warped AdS backgrounds and integrable deformations of spin chains,''
  JHEP {\bf 1305} (2013) 146
  [arXiv:1304.1286 [hep-th]].

\bibitem{magnon} 
  M.~Khouchen and J.~Kluson,
  ``Giant Magnon on Deformed AdS$_3\times$S$^3$\,,''
    Phys.\ Rev.\ D {\bf 90} (2014) 066001
  [arXiv:1405.5017 [hep-th]];
  C.~Ahn and P.~Bozhilov,
  ``Finite-size giant magnons on
$\eta$-deformed AdS$_5 \times $S$^5$,''
   Phys.\ Lett.\ B {\bf 737} (2014) 293
  [arXiv:1406.0628 [hep-th]];
  A.~Banerjee and K.~L.~Panigrahi, 
  ``On the Rotating and Oscillating strings in (AdS$_3\times $S$^3$)$_{\varkappa}$,''
  JHEP {\bf 1409} (2014) 048
  [arXiv:1406.3642 [hep-th]];
 P.~Bozhilov,
  ``Some three-point correlation functions in the eta-deformed $AdS_5\times S^5$,''
  arXiv:1502.00610 [hep-th];   
  A.~Banerjee, S.~Bhattacharya and K.~L.~Panigrahi,
  ``Spiky strings in $\varkappa$-deformed $AdS$,''
  JHEP {\bf 1506} (2015) 057
  [arXiv:1503.07447 [hep-th]]; 
  M.~Khouchen and J.~Kluson,
  ``D-brane on Deformed  $AdS_3\times S^3$,'' 
  JHEP {\bf 1508} (2015) 046
  [arXiv:1505.04946 [hep-th]].

\bibitem{AM-NR} 
  G.~Arutyunov and D.~Medina-Rincon,
  ``Deformed Neumann model from spinning strings on (AdS$_5 \times $S$^5)_\eta$,''
  JHEP {\bf 1410} (2014) 50
  [arXiv:1406.2536 [hep-th]].

\bibitem{Kame-coord}
  T.~Kameyama and K.~Yoshida,
  ``A new coordinate system for $q$-deformed AdS$_5\times$S$^5$ and classical string solutions,''
  J.\ Phys.\ A {\bf 48} (2015) 7,  075401
  [arXiv:1408.2189 [hep-th]].

\bibitem{min-surf}
  T.~Kameyama and K.~Yoshida,
  `` Minimal surfaces in $q$-deformed AdS$_5 \times $S$^5$ string with  Poincare coordinates,''
  J.\ Phys.\ A {\bf 48} (2015) 24,  245401
  [arXiv:1410.5544 [hep-th]]; 
  N.~Bai, H.~H.~Chen and J.~B.~Wu,
  ``Holographic Cusped Wilson loops in $q$-deformed $AdS_5\times S^5$ Spacetime,''
  arXiv:1412.8156 [hep-th].

\bibitem{bi-YB}
  B.~Hoare,
  ``Towards a two-parameter q-deformation of AdS$_3 \times S^3 \times M^4$ superstrings,''
  Nucl.\ Phys.\ B {\bf 891} (2015) 259
  [arXiv:1411.1266 [hep-th]].

\bibitem{LRT}
  O.~Lunin, R.~Roiban and A.~A.~Tseytlin,
  ``Supergravity backgrounds for deformations of AdS$_{n} \times S^n$ supercoset string models,''
  Nucl.\ Phys.\ B {\bf 891} (2015) 106
  [arXiv:1411.1066 [hep-th]].

\bibitem{HT}
  B.~Hoare and A.~A.~Tseytlin,
  ``Type IIB supergravity solution for the T-dual of the eta-deformed $AdS_5 \times S^5$ superstring,''
  arXiv:1508.01150 [hep-th].

\bibitem{Sfetsos}
  K.~Sfetsos,
  ``Integrable interpolations: From exact CFTs to non-Abelian T-duals,''
  Nucl.\ Phys.\ B {\bf 880} (2014) 225
  [arXiv:1312.4560 [hep-th]];
  K.~Sfetsos and D.~C.~Thompson,
  ``Spacetimes for $\lambda$-deformations,''
  JHEP {\bf 1412} (2014) 164
  [arXiv:1410.1886 [hep-th]]; 
 S.~Demulder, K.~Sfetsos and D.~C.~Thompson,
  ``Integrable $\lambda$-deformations: Squashing Coset CFTs and $AdS_5\times S^5$,'' 
  JHEP {\bf 07} (2015) 019
  [arXiv:1504.02781 [hep-th]].
  K.~Sfetsos, K.~Siampos and D.~C.~Thompson, 
  ``Generalised integrable $\lambda$- and $\eta$-deformations and their relation,''
  Nucl.\ Phys.\ B {\bf 899} (2015) 489
  [arXiv:1506.05784 [hep-th]].

\bibitem{Hollowood}
  T.~J.~Hollowood, J.~L.~Miramontes and D.~M.~Schmidtt,
  ``Integrable Deformations of Strings on Symmetric Spaces,''
  JHEP {\bf 1411} (2014) 009
  [arXiv:1407.2840 [hep-th]];
   T.~J.~Hollowood, J.~L.~Miramontes and D.~M.~Schmidtt,
  ``An Integrable Deformation of the $AdS_5\times S^5$ Superstring,''
  J.\ Phys.\ A {\bf 47} (2014) 49,  495402
  [arXiv:1409.1538 [hep-th]]; 
  T.~J.~Hollowood, J.~L.~Miramontes and D.~M.~Schmidtt,
  ``S-Matrices and Quantum Group Symmetry of k-Deformed Sigma Models,''
  arXiv:1506.06601 [hep-th]; 
C.~Appadu and T.~J.~Hollowood,
  ``Beta Function of k Deformed $AdS_5\times S^5$ String Theory,''
 arXiv:1507.05420 [hep-th].

\bibitem{Vicedo}
  B.~Vicedo,
  ``Deformed integrable $\sigma$-models, classical $R$-matrices 
and classical exchange algebra on Drinfel'd doubles,''
  arXiv:1504.06303 [hep-th].

\bibitem{Hoare-Tseytlin}
  B.~Hoare and A.~A.~Tseytlin,
  ``On integrable deformations of superstring sigma models related to $AdS_n \times S^n$ supercosets,''
  Nucl.\ Phys.\ B {\bf 897} (2015) 448
  [arXiv:1504.07213 [hep-th]].

\bibitem{Klimcik-lambda}
C.~Klimcik,
  ``$\eta$ and $\lambda$ deformations as ${\cal E}$-models,''
  arXiv:1508.05832 [hep-th].





\end{thebibliography}
\end{document}